\documentclass[final]{cvpr}

\usepackage{times}
\usepackage{epsfig}
\usepackage{graphicx}
\usepackage{amsmath}
\usepackage{amssymb}

\usepackage[pagebackref=true,breaklinks=true,colorlinks,bookmarks=false]{hyperref}

\newif\ifarXiv
\arXivtrue 

\ifarXiv
\usepackage[title]{appendix}
\else
\fi

\def\cvprPaperID{8456} 
\def\confYear{CVPR 2021}

\DeclareMathOperator*{\argmin}{arg\,min}

\usepackage{soul}

\newcommand{\textapprox}{\raisebox{0.5ex}{\texttildelow}}
\newcommand{\myparagraph}[1]{\vspace{-10pt}\paragraph{#1}}
\begin{document}

\title{What's in the Image?\\Explorable Decoding of Compressed Images}

\author{Yuval Bahat and Tomer Michaeli\\
Technion - Israel Institute of Technology, Haifa, Israel\\
{\tt\small \{yuval.bahat@campus,tomer.m@ee\}.technion.ac.il}}

\maketitle
\ifarXiv\else
\thispagestyle{empty}
\fi
\begin{abstract}
    The ever-growing amounts of visual contents captured on a daily basis necessitate the use of lossy compression methods in order to save storage space and transmission bandwidth. While extensive research efforts are devoted to improving compression techniques, every method inevitably discards information. 
    Especially at low bit rates, this information often corresponds to semantically meaningful visual cues, so that decompression involves significant ambiguity. In spite of this fact, existing decompression algorithms typically produce only a single output, and do not allow the viewer to explore the set of images that map to the given compressed code. 
    In this work we propose the first image decompression method to facilitate user-exploration of the diverse set of natural images that could have given rise to the compressed input code, thus granting users the ability to determine what could and what could not have been there in the original scene.
    Specifically, we develop a novel deep-network based decoder architecture for the ubiquitous JPEG standard, which allows traversing the set of decompressed images that are consistent with the compressed
    JPEG file.
    To allow for simple user interaction, we develop a graphical user interface comprising several intuitive exploration tools, including an automatic tool for examining specific solutions of interest.
    We exemplify our framework on graphical, medical and forensic use cases, demonstrating its wide range of potential applications.
\end{abstract}
\section{Introduction}\label{sec:intro}
\begin{figure*}[t]
\centering
\includegraphics[width=\textwidth]{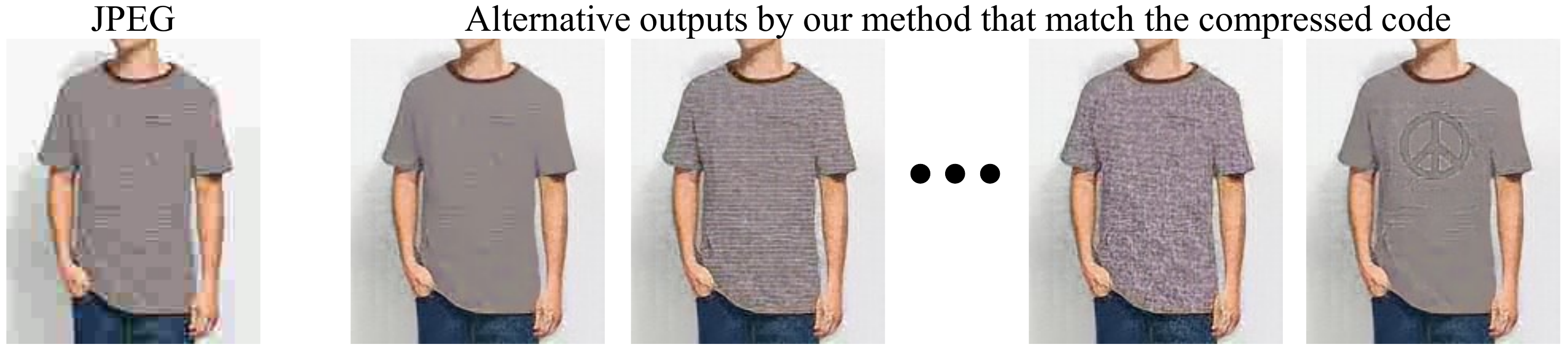}
\vspace{-10pt}\caption{\label{fig:teaser}\textbf{Ambiguity in JPEG decompression.} A compressed JPEG file can correspond to numerous different plausibly looking images. These can vary in color, texture, and other structures that encode important semantic information. Since multiple images map to the same JPEG code, any decoder that outputs only a single reconstruction, fails to convey to the viewer the ambiguity regarding the encoded image.}
\end{figure*}

With surveillance systems so widely used and social networks ever more popular, the constant growth in the capacity of daily captured visual data necessitates using lossy compression algorithms (\eg JPEG, H.264), that discard part of the recorded information in order to reduce storage space and transmission bandwidth. Over the years, extensive research has been devoted to improving compression techniques, whether by designing better decoders for existing encoders, or by devising new compression-decompression (CODEC) pairs, that enable
higher perceptual quality even at low bit-rates.
However, in any lossy compression method, the decoder faces inevitable ambiguity.
This ambiguity is particularly severe at low bit-rates, which are becoming more prevalent with the ability to maintain perceptual quality at extreme compression ratios \cite{agustsson2019end2end_perception_gan}.
This is exemplified in Fig.~\ref{fig:teaser} in the context of the JPEG standard. Low bit-rate compression may prevent the discrimination between different animals, or the correct identification of a shirt pattern, a barcode, or text. Yet, despite this inherent ambiguity, existing decoders do not allow the user to explore the abundance of plausible images that could have been the source of a given compressed code.

\begin{figure*}[t]
\centering
\includegraphics[width=\textwidth]{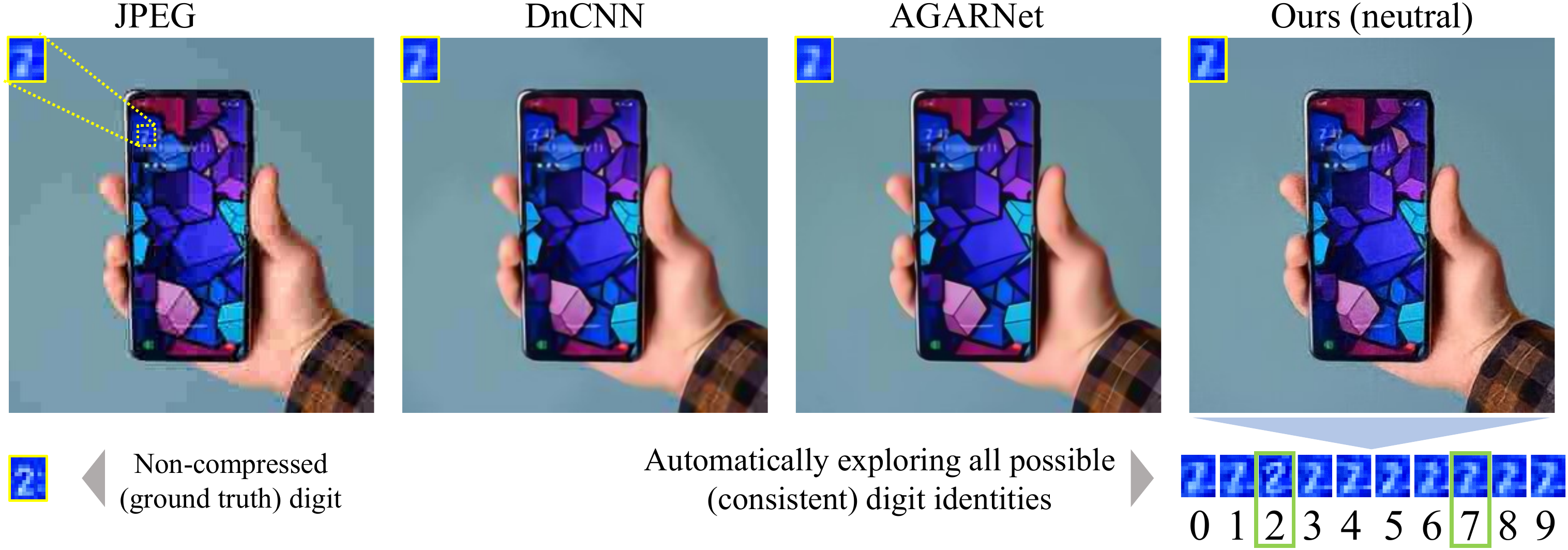}
\vspace{-10pt}\caption{\label{fig:auto_exploration}\textbf{Automatic exploration.} Upon marking an ambiguous character in the image, our GUI harnesses a pre-trained digit classifier to propose optional (consistent) reconstructions corresponding to the possible digits $0-9$ (see details in Sec.~\ref{sec:editing_tools}). This feature is valuable in many use cases (\eg forensic), as it can assist in both revealing and resolving decompression ambiguities; although pre-exploration decoding of the hour digit (yellow rectangle) by all methods (top row) may suggest it is $7$, our automatic exploration tool produces perceptually plausible decodings as both $7$ and $2$ (green rectangles), thus uncovering the hidden ambiguity and even preventing false identification, as the correct hour digit (in the pre-compressed image, bottom left) was indeed $2$.}
\end{figure*}
Recently, there has been growing research focus on models that can produce diverse outputs for any given input, for image synthesis \cite{zhu2017multimodal_image2image,chen2018diverse_sketch2image,lee2018diverse_image2image}, as well as for image restoration tasks, \eg denoising \cite{prakash2020divnoising}, compression artifact reduction \cite{guo2017one2many} and super-resolution \cite{bahat2019explorable_sr,buhler2020explorative_sr,lugmayr2020srflow}. The latter group of works took another step, and also allowed users to interactively traverse the space of high-resolution images that correspond to a given low-resolution input. In this paper, we propose the first method to allow users to explore the space of natural images that corresponds to a compressed image code. We specifically focus on the ubiquitous JPEG standard, though our approach can be readily extended to other image and video compression formats.

A key component of our method is a novel JPEG decompression network architecture, which predicts the quantization errors of the DCT coefficients and is thus guaranteed to produce outputs that are consistent with the compressed code. This property is crucial for enabling reliable exploration and examining what could and what could not have been present in the underlying scene. Our scheme has a control input signal that can be used to manipulate the output. This, together with adversarial training, allows our decoder to generate diverse photo-realistic outputs for any given compressed input code. 

\begin{figure*}[t]
\centering
\includegraphics[width=\textwidth]{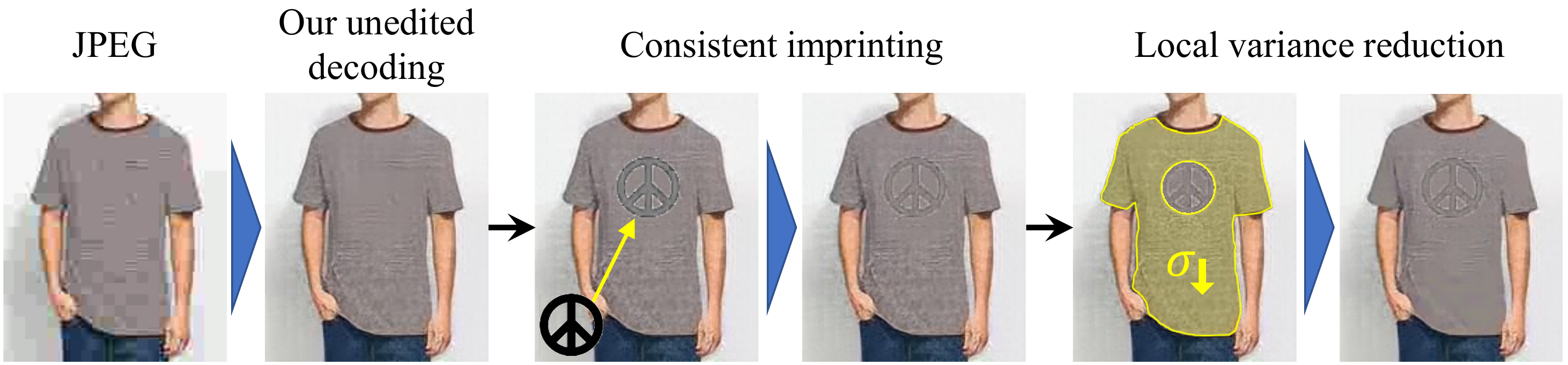}
\vspace{-10pt}\caption{\label{fig:editing_teaser}\textbf{Example exploration process.} Our GUI enables the user to explore the enforcement of various properties on any selected region within the image. Unlike existing editing methods that only impose photo-realism, ours seeks to conform to the user's edits while also restricting the output to be \emph{perfectly consistent} with the compressed code.
\vspace{-5pt}
}
\end{figure*}

We couple our network with a graphical user interface (GUI), which lets the user interactively explore the space of consistent and perceptually plausible reconstructions. The user can attempt to enforce contents in certain regions of the decompressed image using various tools (see \eg Fig.~\ref{fig:editing_teaser}). Those trigger an optimization problem that determines the control signal best satisfying the user's constraints. Particularly, our work is the first to facilitate \emph{automatic} user exploration, by harnessing pre-trained designated classifiers, \eg to assess which digits are likely to correspond to a compressed image of a digital clock (see Fig.~\ref{fig:auto_exploration}).

\begin{figure}[t]
\centering
\includegraphics[width=\columnwidth]{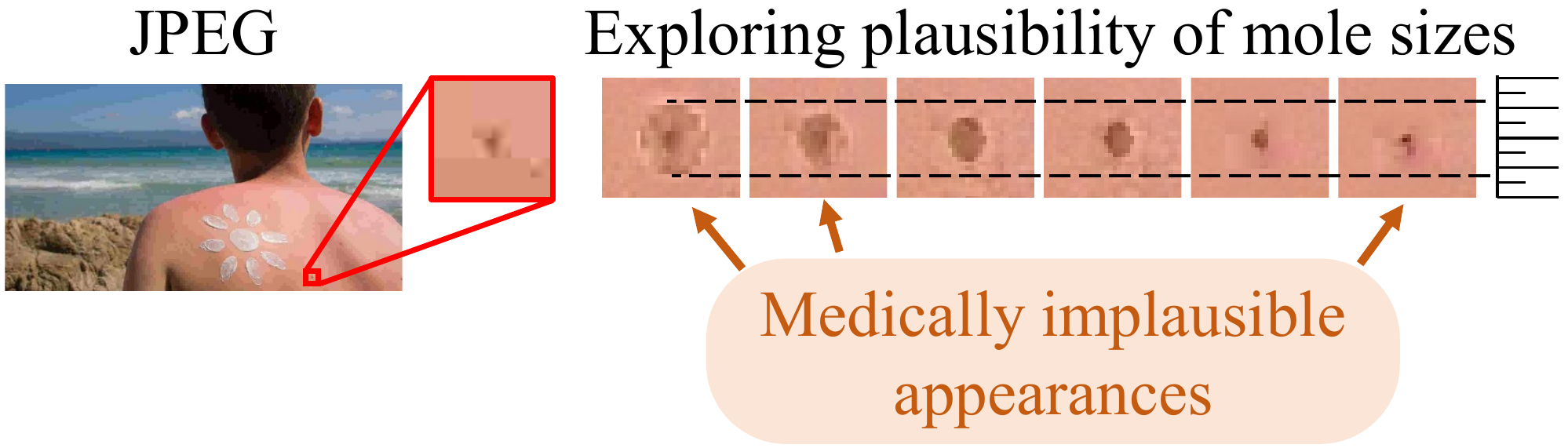}
\vspace{-10pt}\caption{\label{fig:applications}
\textbf{Medical application example.} A dermatologist examining a suspected mole on a new patient may turn to existing patient photos containing this mole, to study its development. As the mole appearance in such images may often be degraded due to compression, our method can assist diagnosis by allowing exploration of the range of possible mole shapes and sizes. Please see corresponding editing processes in supplementary.
\vspace{-10pt}
}
\end{figure}

Our explorable JPEG decoding approach is of wide applicability. Potential use cases range from allowing a user to restore lost information based on prior knowledge they may have about the captured image, through correcting unsatisfying decompression outputs (demonstrated in Fig.~\ref{fig:dunes}), to situations where a user wants to test specific hypotheses regarding the original image. The latter setting is particularly important in forensic image analysis and in medical image analysis, as exemplified in Figs.~\ref{fig:auto_exploration} and~\ref{fig:applications}, respectively.
\section{Related Work}\label{sec:related}

\myparagraph{Diverse and explorable image restoration}
Recently, there is growing interest in image restoration methods that can generate a diverse set of reconstructions for any given input. Prakash \etal \cite{prakash2020divnoising} proposed to use a variational autoencoder for diverse denoising. Guo \etal \cite{guo2017one2many} addressed diverse decompression, allowing users to choose between different decompressed outputs for any input compressed image. In the context of super-resolution, the GAN-based PULSE method \cite{menon2020pulse} can produce diverse outputs by using different latent input initializations, while the methods in \cite{bahat2019explorable_sr,buhler2020explorative_sr,lugmayr2020srflow} were the first to allow user manipulation of their super-resolved outputs. Note that among these methods, only \cite{bahat2019explorable_sr} guarantees the consistency of all its outputs with the low-resolution input, which is a crucial property for reliable exploration, \eg when a user is interested in assessing the plausibility of a specific solution of interest.

Though we borrow some ideas and mechanisms from explorable super-resolution \cite{bahat2019explorable_sr}, this work is the first to discuss the need and propose a framework for performing \emph{explorable image decompression}, which is a fundamentally more challenging task.
While in super-resolution, the set of consistent solutions is a linear subspace that has zero volume in the space of all high-resolution images (just like a 2D plane has zero volume within 3D space),
image compression involves quantization and so induces a set of consistent reconstructions with nonzero volume (just like a cube occupies nonzero volume in 3D space). We therefore introduce various novel mechanisms, including a fundamentally different consistency enforcing architecture, novel editing tools tailored for image decompression, and an \emph{automatic} exploration tool (see Fig. 2), that is invaluable for many applications (e.g. forensics).
Though ours is the first JPEG decompression method aiming for perceptual quality that is guaranteed to generate consistent reconstructions, we note that Sun \etal \cite{sun2020dct_consistent} proposed a consistent decompression scheme, but which is aimed at minimizing distortion rather than maximizing photo-realism (thus outputting the mean of the plausible explanations to the input).

\myparagraph{Improved JPEG decompression} 
Many works proposed improved decompression techniques for existing compression standards \cite{lee2004regression_predict,chen2001adaptive_dct_smoothing,liu2015dual_domain_sparsity,dar2016admm_linearization,liu2016random_walk_non_dl,fu2019ar_pixel_dilated,ororbia2019ar_rnn,zhang2017dncnn,zini2020rrdb_autoencoder,kim2020per_pixel_qf_est}. 
Specifically for JPEG, classical artifact reduction (AR) methods \cite{lee2004regression_predict,chen2001adaptive_dct_smoothing,liu2015dual_domain_sparsity} attempted different heuristics, like smoothing DCT coefficients \cite{chen2001adaptive_dct_smoothing} or relying on natural image priors like sparsity, in both DCT and pixel domains \cite{liu2015dual_domain_sparsity}. Deep convolutional AR networks (first proposed by Dong \etal \cite{dong2015arcnn}) learn to minimize a reconstruction error with respect to ground truth reference images, and operate either in the pixel \cite{dong2015arcnn,svoboda2016deeper_with_res,zhang2017dncnn,zini2020rrdb_autoencoder}, DCT \cite{Yoo_2018dct_cross_entropy,sun2020dct_consistent} or both domains \cite{wang2016d3,guo2016ddcn,guo2017one2many,zhang2018dmcnn,kim2020per_pixel_qf_est}. 
Some recent AR methods \cite{galteri2017single_qf_gan,galteri2019multiple_qf_gans,kim2019constrained_gan} use a generative adversarial network (GAN) scheme \cite{goodfellow2014gan} for encouraging more photo-realistic results, which we too employ in our framework. 
Our design (like \cite{zhang2017dncnn,zini2020rrdb_autoencoder,kim2020per_pixel_qf_est}) is oblivious to the quality factor (QF) parameter, and can therefore handle a wide range of compression levels. 
In contrast, other methods are trained for a fixed QF setting, which is problematic not only because it requires training a different model for each QF, but also since QF by itself is an ambiguous parameter, as its conversion into compression level varies across implementations.

\section{Our Consistent Decoding Model}\label{sec:our_model}
To enable exploration of our decompression model's outputs, we need to verify they are both perceptually plausible and consistent with the given compressed code. To satisfy the first requirement, we adopt the common practice of utilizing an adversarial loss, which penalizes for deviations from the statistics of natural images. To satisfy the consistency requirement, we introduce a novel network design that is specifically tailored for the JPEG format.
The JPEG encoding scheme works in the $Y-Cb-Cr$ color space and uses separate pipelines for the luminance ($Y$) and chrominance ($Cb$, $Cr$) channels.
Our model supports color images, however for clarity we start by describing the simpler case of gray-scale images. The non-trivial treatment of color is deferred to Sec.~\ref{sec:color}.
We begin with a brief description of the relevant components in the JPEG compression pipeline, before describing our network design.

\begin{figure*}[t]
\centering
\includegraphics[width=\textwidth]{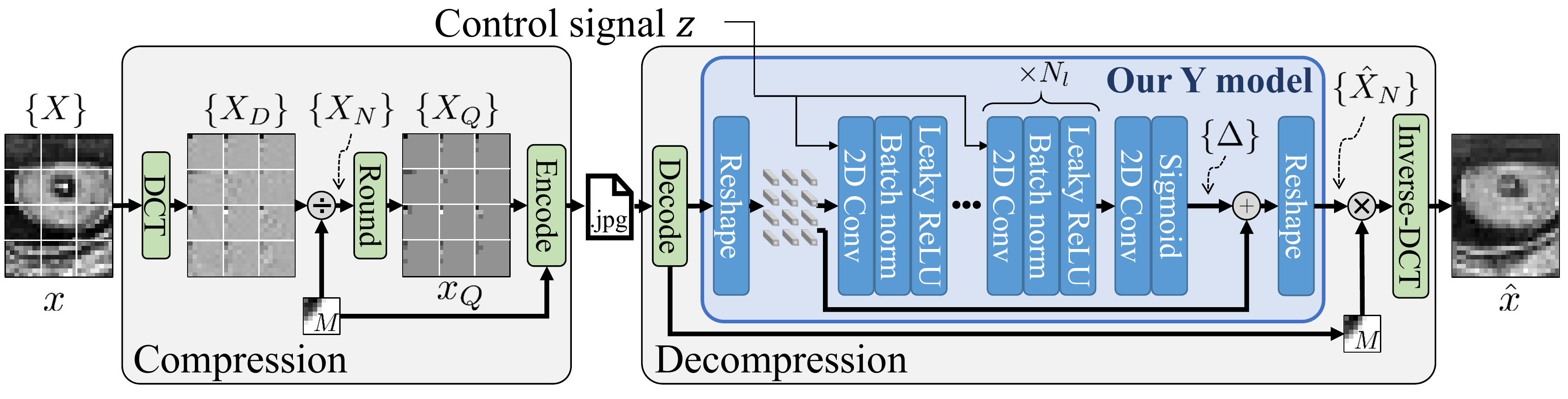}
\caption{\label{fig:framework}\textbf{Gray-scale JPEG compression scheme and our model.} Our network (inside the blue rectangle) is incorporated into the JPEG decompression pipeline in a way that guarantees the consistency of its outputs with the JPEG file content, while yielding a significant perceptual quality gain compared to images decompressed using the standard pipeline. An additional input signal $z$ is incorporated to allow manipulating the output. Please refer to the description in Sec.~\ref{sec:our_model}.}
\end{figure*}
\subsection{JPEG compression}
The encoding process is shown at the left hand side of Fig.~\ref{fig:framework}. It starts by dividing the input image $x$, which is assumed to be of size\footnote{We assume integer $m$ and $n$ only for simplicity. Arbitrary image sizes are also supported.} $8m\times 8 n$, into an $m\times n$ array of
non-overlapping $8\times8$ blocks. For each $8\times 8$ block $X$, the encoder computes its DCT, $X_D=\text{DCT}(X)$, and divides it element-wise by a pre-defined matrix $M \in {\mathbb Z}^{8\times8}$ to obtain a block of normalized DCT coefficients $X_N=X_D\oslash M$. Finally, the entries of $X_N$ are rounded to yield a block of quantized coefficients $X_Q=\text{round}(X_N)$.
The blocks $\{X_Q\}$ are stored into the JPEG file alongside matrix $M$ using some additional \emph{lossless} processing steps.
Note that the matrix $M$ comprises the per-coefficient quantization intervals, determined as a function of the scalar QF parameter. However, this function varies between JPEG implementations, and therefore the QF itself is ambiguous and insufficient for extracting the image.

\subsection{Our decoder design}\label{sec:decoder_design}
Our decoder network is shown at the right hand side of Fig.~\ref{fig:framework}. Our network operates in the DCT domain.
Namely, for each $8\times 8$ block $X_Q$ extracted from the file, our network outputs an estimate ${\hat X}_D$ of the corresponding block $X_D$. The decoded image is then constructed by applying inverse DCT on the predicted DCT blocks $\{{\hat X}_D\}$. To predict $X_D$, we first generate a prediction ${\hat X}_N$ of the normalized coefficients $X_N$, and then multiply it element-wise by $M$, so that ${\hat X}_D$ = ${\hat X}_N \odot M$.
Since information loss during image encoding is only due to the rounding step, we consider a reconstructed block ${\hat X}_N$ to be consistent with the quantized block $X_Q$ when it satisfies ${\hat X}_N=X_Q+\Delta$, with an $8\times 8$ matrix $\Delta$ whose entries are all in $[-0.5,0.5)$. We therefore design our network to predict this $\Delta$ for each block, and we confine its entries to the valid range using a shifted Sigmoid function. This process guarantees that the decoded image is perfectly consistent with the compressed input code.

Predicting the residual $\Delta$ for each block $X_Q$ is done as follows. We arrange the blocks $\{X_Q\}$ to occupy the channel dimension of an $m \times n \times 64$ tensor $x_Q$, so that each block retains its relative spatial position w.r.t.~the other blocks in the image.
We then input this tensor to a network comprising $N_\ell$ layers of the form \mbox{2-D convolution $\rightarrow$ batch normalization $\rightarrow$ leaky ReLU}, followed by an additional 2-D convolution and a Sigmoid. All convolution kernels are $3\times 3$. The last convolution layer outputs $64$ channels, which correspond to the residual blocks $\{\Delta\}$. Compared to operating in the pixel domain, the receptive field of this architecture is $8\times$ larger in each axis, thus allowing it to exploit larger scale cues.

An important distinction of our network is the ability to manipulate its output, which facilitates our goal of performing explorable image decompression. This is enabled by incorporating a control input signal, which we feed to the network in addition to the quantized input $x_Q$. We define our control signal $z\in{\mathbb R}^{m\times n\times 64}$ to have the same dimensions as $x_Q$, so as to allow intricate editing abilities.
Following the practice in \cite{bahat2019explorable_sr}, we concatenate $z$ to the input of each of the $N_\ell$ layers of our network, to promote faster training.

\begin{figure*}[t]
\centering
\includegraphics[width=\textwidth, trim=0cm 0.3cm 0cm 0.3cm, clip]{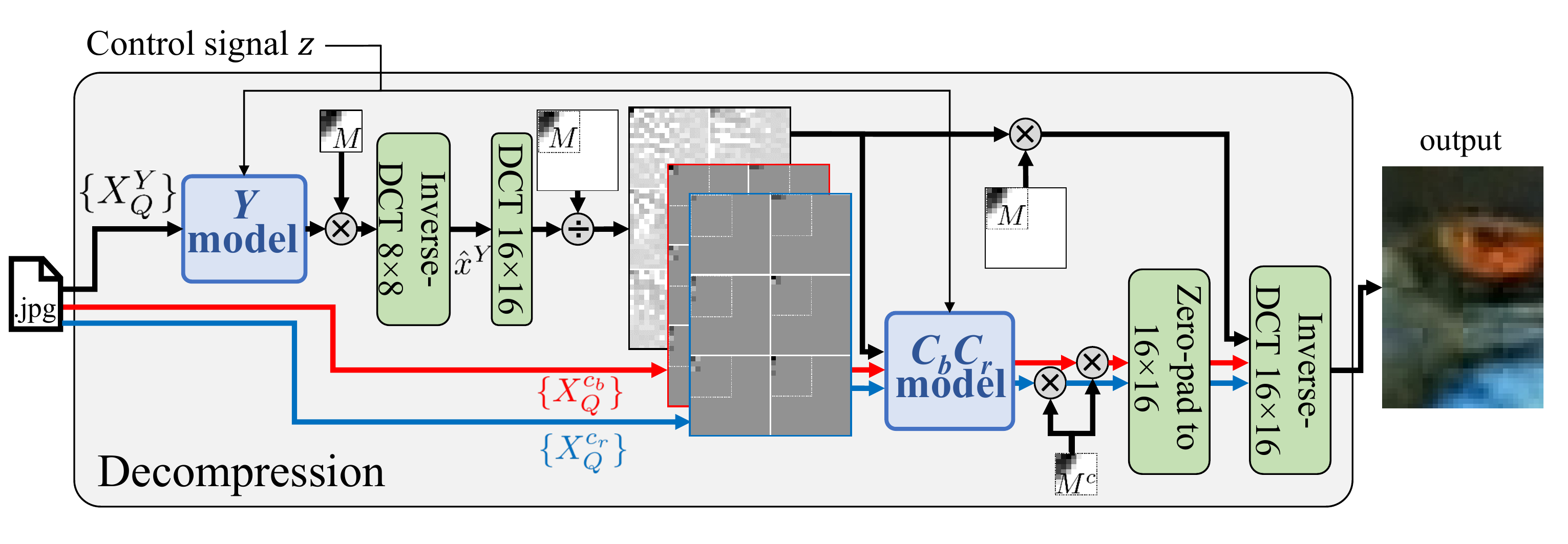}
\caption{\label{fig:color_framework}\textbf{Our full image decompression scheme.} We employ separate models (blue rectangles) to compensate for the quantization errors of the luminance and chroma channels. Both models share the same internal design depicted in Fig.~\ref{fig:framework}, and receive the same control signal $z$, to allow coordinated editing. Please refer to the full description in Sec.~\ref{sec:color}.
\vspace{-10pt}
}
\end{figure*}
We train our model following the general procedure of~\cite{bahat2019explorable_sr}. As an initialization step, we train it to minimize the $L_1$ distance between ground truth uncompressed training images, and the corresponding outputs of our network, while randomly drawing the QF parameter for each image. Once initialized, training continues without utilizing any full-reference loss terms like $L_1$ or VGG, which is made possible thanks to the inherent consistency guarantee of our design. These full-reference loss terms are known to bias the output towards the (overly-smoothed) average of all possible explanations to the given compressed code, and are thus not optimal for the purpose of exploration. We instead minimize the following weighted combination of loss terms to guide our model:
\begin{equation}
    \label{eq:total_loss}
    {\cal L}_\text{Adv}+\lambda_\text{Range}{\cal L}_\text{Range}
    +\lambda_\text{Map}{\cal L}_\text{Map}.
\end{equation}
Here, ${\cal L}_\text{Adv}$ is an adversarial loss, which encourages the reconstructed coefficient blocks ${\hat X}_D$ to follow the statistics of their natural image counterparts. In particular, we employ a Wasserstein GAN loss with spectral normalization \cite{miyato2018spectral_normalization} and gradient penalty \cite{gulrajani2017wgan_gp}, and use the same model architecture for both generator and critic (except for substituting batch normalization with layer normalization in the latter), following the recommendations in \cite{gulrajani2017wgan_gp}.
The second loss term, ${\cal L}_\text{Range}$, helps prevent model divergence by penalizing for pixel values outside $[16,235]$, which is the valid range of luminance values in the $Y-Cb-Cr$ color space. We use ${\cal L}_\text{Range}=\tfrac{1}{k}\|\hat{x}-\text{clip}_{[16,235]}\{\hat{x}\}\|_1$, where $k=64\cdot m\cdot n$ is the number of pixels in the image. 

The last loss term in~\eqref{eq:total_loss} is associated with the control signal input $z$, which at exploration (test) time should allow traversing the space of plausible consistent decompressed images.
We therefore use ${\cal L}_\text{Map}=\min_z\|\psi(x_Q,z)-x\|_1$ to (i) prevent our network $\psi$ from ignoring its input $z$, as well as to (ii) guarantee our network can produce each ground truth natural image $x$ in our training set using \emph{some} $z$. This is in addition to the adversarial loss ${\cal L}_{\text{Adv}}$, which encourages the network's output corresponding to \emph{any} input $z$ to be perceptually plausible.
Within each training step, we first compute $z^*=\argmin_z\{{\cal L}_\text{Map}\}$ using $10$ iterations, and then use the fixed $z^*$ for the minimization of all loss terms in~\eqref{eq:total_loss}.

\subsection{Training details}
We train our model with the Adam optimizer on 1.15M images from the ImageNet training set \cite{russakovsky2015imagenet}, using batches of $16$ images.
The initialization and consecutive training phases last 6 and 12 epochs and employ learning rates of $10^{-4}$ and $10^{-5}$, respectively.
Batches in the latter phase are fed twice, once with a random $z$ and once with the optimal $z^*$ minimizing ${\cal L}_\text{Map}$ (see details in the Supp.). We set $\lambda_\text{Range}$ and $\lambda_\text{Map}$ to $200$ and $0.1$, respectively.
To create training input codes, we compress the GT training images utilizing a quantization interval matrix 
$M=50\cdot Q_{\text{baseline}}/\text{QF}$, where $Q_{\text{baseline}}$ is the example baseline table in the JPEG standard \cite{wallace1992jpeg} and QF is independently sampled from a uniform distribution over $[5,49]$ for each image\footnote{
QFs in the range $[50,100]$ induce lower data loss, and are thus less interesting for explorable decoding.
The effect of control signal $z$ in such high QFs is only minor, thanks to our framework's consistency guarantee.
}. We use $N_\ell=10$ layers for both the generator and the critic models, using $320$ output channels for all convolution operations but the last. We employ a conditional critic, by
concatenating the generator's input $x_Q$ to our critic's input, as we find it to accelerate training convergence.

\section{Handling Color Channels}\label{sec:color}
\begin{figure*}[ht]
\centering
\includegraphics[width=\textwidth]{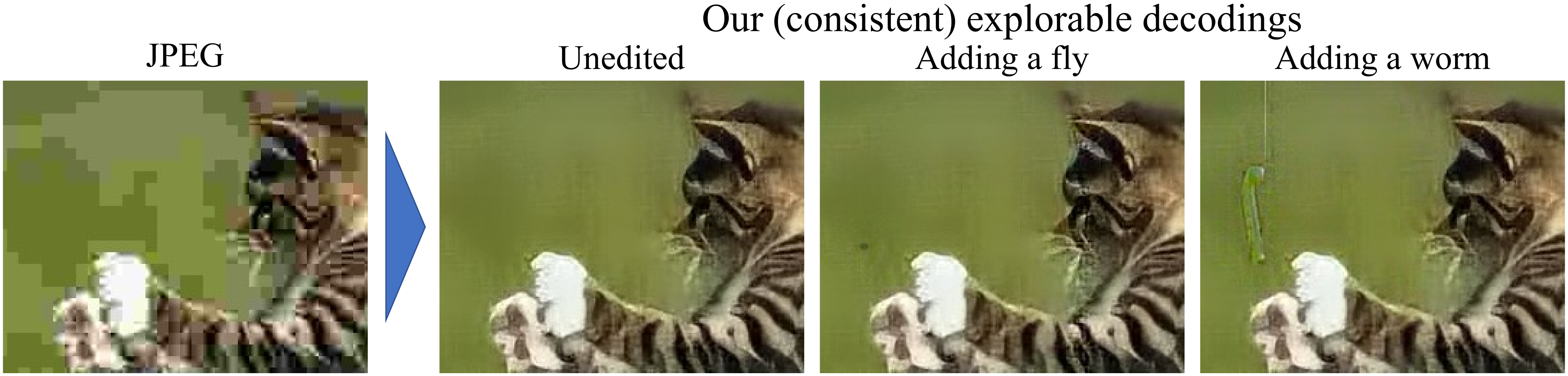}
\vspace{-10pt}\caption{\label{fig:kitten_and_fly}\textbf{Exploring plausible explanations.} Artifacts in the given compressed image (left) are removed by our method (middle-left) prior to any exploration. We can then use our GUI to produce different explanations to the kitten's attention, by imprinting, \eg a tiny fly or a worm onto the unedited image. These alternative reconstructions perfectly match the JPEG code when re-compressed. Please refer to the Supp.~for a visualization of the control signals $z$ corresponding to each output.
}
\end{figure*}

\begin{figure*}[ht]
\centering
\includegraphics[width=\textwidth]{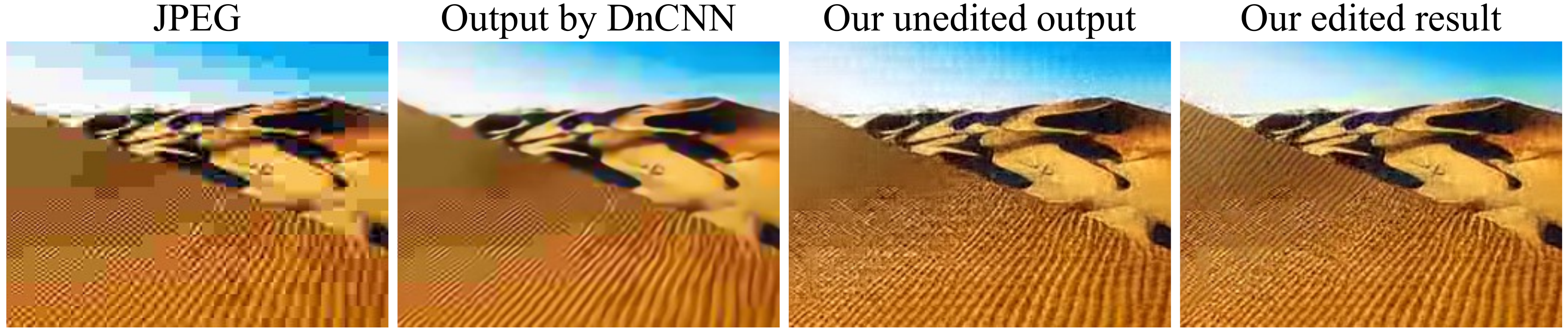}
\vspace{-10pt}\caption{\label{fig:dunes}\textbf{Correcting unpleasing decompression.} Existing artifact removal methods like DnCNN \cite{zhang2017dncnn} (middle-left), are often able to ameliorate compressed images (left), but do not allow editing their output. In contrast, outputs by our method (middle-right) can be edited by a user to yield superior results (right), which are guaranteed to match the input JPEG code if re-compressed.
\vspace{-5pt}
}
\end{figure*}
Let us denote the channels of a color image $x$ by $x^{Y}$, $x^{Cb}$, and $x^{Cr}$. The chroma channels ($Cb$ and $Cr$) of natural images tend to contain mostly low frequency content. The JPEG format exploits this fact, by allowing to subsample those channels in the pixel space. The subsampled channels are then divided into $8\times 8$ blocks, whose DCT coefficients are quantized using matrix $M^c$, similarly to the luminance channel. These quantized blocks, denoted by $\{X_Q^{Cb}\}$ and $\{X_Q^{Cr}\}$, are stored in the color JPEG file alongside their luminance channel counterparts $\{X^Y_Q\}$. This process results in lost chroma information, which like its luminance counterpart, may correspond to semantically meaningful visual cues. Our framework allows exploring both the luminance and the chrominance of the image,
as depicted in Fig.~\ref{fig:color_framework}.

Here we use the most aggressive ``4:2:0'' subsampling configuration of JPEG, corresponding to subsampling the chroma channels by a factor of $2$ in both axes. We reconstruct the chroma channels using an additional network, which handles the chroma information loss due to quantization. While we can use the same process employed in the luminance case to handle the chroma quantization, accounting for subsampling requires some modifications. Before elaborating on these modifications, we begin by briefly describing the relevant steps in the JPEG chroma pipeline.

\subsection{Alternative modeling of chroma subsampling}\label{sec:subsampling_remodeling}
Ideally, we would have liked to inform the chroma reconstruction network of the luminance information, by concatenating the luminance and chroma codes. However, this is impractical due to the spatial dimension mismatch resulting from the chroma subsampling. To overcome this hurdle, we remodel the above described subsampling pipeline using an approximated pipeline  as follows.

In natural image chroma channels, almost all content is concentrated at the low frequencies (\eg, in the BSD-100 dataset \cite{Martin2001bsd}, an average of $99.99998\%$ of the energy in each $16\times 16$ chroma channel DCT block, is concentrated in the upper-left $8\times 8$ sub-block). For such low-frequency signals, the above mentioned subsampling procedure incurs neglieible aliasing. Thus, the $8\times 8$ DCT blocks of the subsampled channels can be assumed (for all practical purposes) to have been constructed by first computing the DCT of each $16 \times 16$ block of the original chroma channels, and then extracting from it only the $8\times 8$ block of coefficients corresponding to the low-frequency content. The rest of the process is modeled without modification. As we show in the Supplementary, the differences between images processed using the actual and approximated pipelines are negligible (\eg the PSNR between the two is $88.9$dB over BSD-100).

\subsection{Modifying our design to support subsampling}
Given a compressed input code, we first reconstruct the luminance channel, as described in Sec.~\ref{sec:our_model}. The reconstructed luminance image $\hat{x}^{Y}$ is then fed into a chroma decoding network together with the quantized chroma blocks from the JPEG file, to obtain the final decoded color image.

Since the quantized $8\times 8$ blocks of the chroma channels in fact correspond to $16\times 16$ blocks of the image, our network operates on $16\times 16$ blocks. Specifically, for the luminance channel $\hat{x}^{Y}$, we compute DCT coefficients for each $16\times 16$ block and reshape them into a tensor with $16^2=256$ channels. The $8\times 8$ chroma blocks stored in the file are zero-padded to be $16 \times 16$ in size (so that the high frequencies are all zeros) and then also reshaped into tensors with $256$ channels (see Fig.~\ref{fig:color_framework}). The luminance tensor is concatenated with the chrominance tensors to form a single tensor with $3\times 256=768$ channels\footnote{In practice, we discard the channels corresponding to the zero-padding, which are all zeros.}. This tensor is then fed into our chroma network, which uses the same architecture described in Sec.~\ref{sec:our_model}, only with $160$ channels in the internal layers. This network yields error estimate blocks $\Delta$ of size $8\times 8$, which are added to the quantized blocks ${X}_Q^{Cb}$ and ${X}_Q^{Cr}$, and multiplied by ${M}^c$. The resulting blocks are zero-padded to $16\times 16$ (setting the high frequencies to zero) and converted back to pixel-space using inverse DCT. These reconstructed chroma channels are then combined with the luminance channel to yield the reconstructed color image. We feed the same control input signal $z$ to both luminance and chroma models, to allow a coordinated editing effect.
\section{Exploration Tools and Use Cases}\label{sec:editing_tools}
\begin{figure*}[th]
\centering
\includegraphics[width=\textwidth]{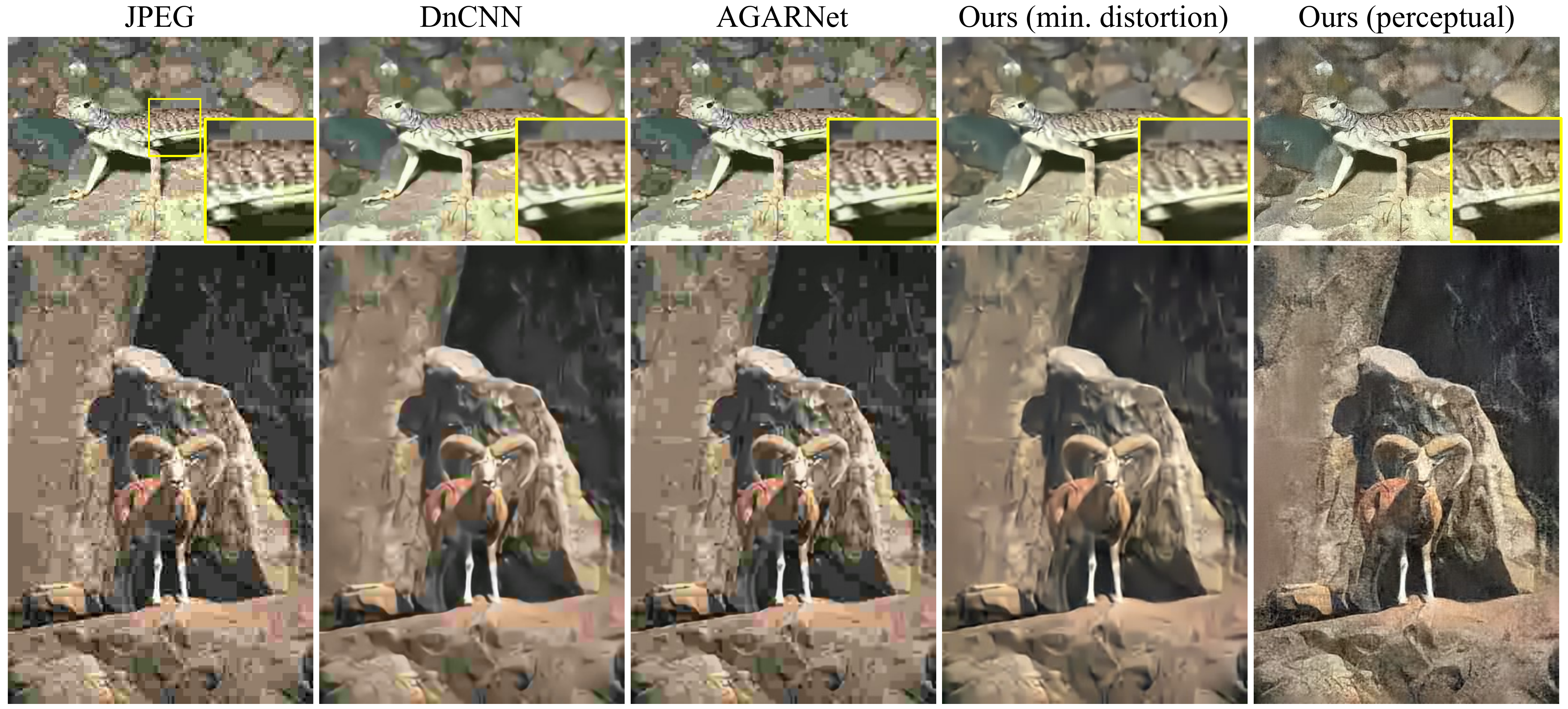}
\vspace{-10pt}\caption{\label{fig:visual_comparison}\textbf{Qualitative comparison.} Our GAN-trained image decompression model is primarily intended to allow consistent user exploration, especially for very low QFs, which induce larger ambiguities. Nonetheless, as demonstrated here on severely compressed images (QF=$5$), even pre-edited  outputs (corresponding to random $z$ inputs) of our model
(right) yield significant perceptual quality gains relative to the JPEG decompression pipeline (left), as well as compared to results by the DnCNN \cite{zhang2017dncnn} \&  AGARNet \cite{kim2020per_pixel_qf_est} AR methods, and a variant of our model trained to minimize distortion (middle columns). Please refer to the supplementary for additional visual comparisons.}
\end{figure*}
\begin{figure*}[th]
\vspace{-5pt}
\centering
{\includegraphics[width=\columnwidth]{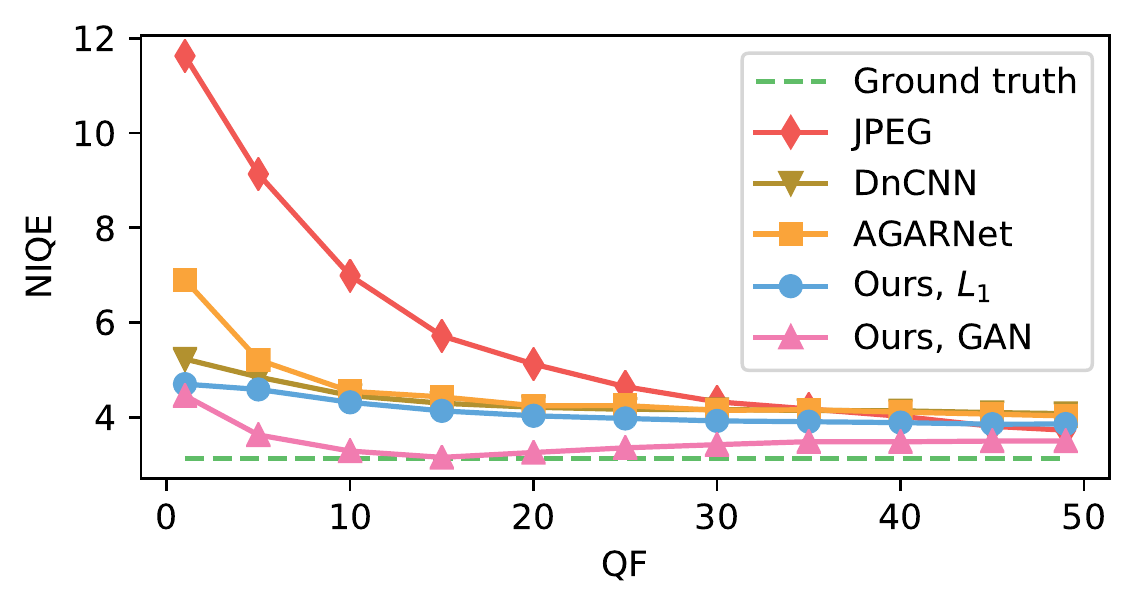}}
\qquad
{\includegraphics[width=\columnwidth]{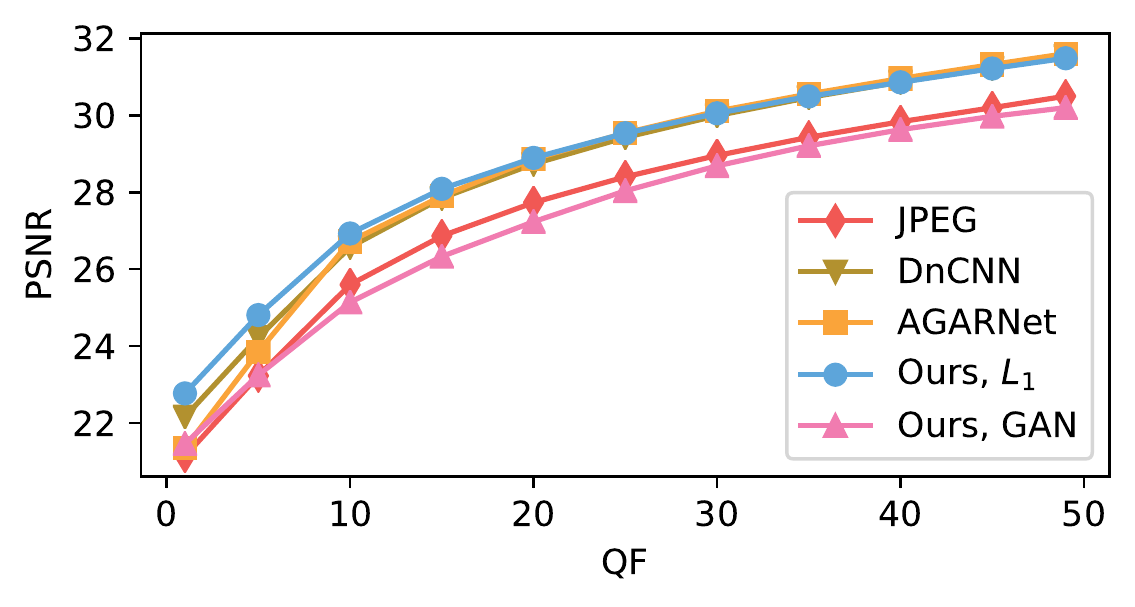}}
\vspace{-17pt}
\caption{\label{fig:quantitative_comparison}\textbf{Quantitative evaluation.} Comparing our exploration model (Ours, GAN), and its 
variant trained to minimize distortion 
(Ours, $L_{1}$), with the DnCNN \cite{zhang2017dncnn} and AGARNet \cite{kim2020per_pixel_qf_est} AR methods on the BSD-100 \cite{Martin2001bsd} dataset, in terms of perceptual quality (left, lower is better) and image distortion (right, higher is better). Please refer to the Supp. for more comparisons and additional details.}
\vspace{-10pt}
\end{figure*}
Having trained both luminance and chroma models, we facilitate user exploration by employing a \emph{graphical user interface} (GUI) comprising different editing tools. Our GUI runs on an NVIDIA GeForce 2080 GPU, and allows interactive exploration in real time.
Specifically, once a compressed image code $x_Q$ is loaded from a JPEG file, a user can manipulate the output of our decoding network, $\hat{x}=\psi(x_Q,z)$, by first marking a region to be edited and then choosing among different available tools. Those enable the user to attempt enforcing various properties on $\hat{x}$. Each editing tool triggers a process of solving $z^*=\argmin_{z} f(\psi(x_Q,z))$ behind the scenes, for some objective function~$f$, which is optimized using the Adam optimizer. The result is a modified output image $\psi(x_Q,z^*)$, which is guaranteed to be consistent with the compressed code $x_Q$ (due to our network's architecture) and to have a natural appearance (due to the parameters of $\phi$ which have been shaped at train time to favor natural outputs). Examples for such images $\hat{x}$ are depicted in Fig.~\ref{fig:kitten_and_fly}.

We introduce a very useful automatic exploration tool, which given an image region, presents optional reconstructions corresponding to each digit $d\in\{0,\dots,9\}$, by utilizing $f(\cdot)=\text{Classifier}_d(\cdot)$, the output of a pre-trained digit classifier \cite{goodfellow2013svhn_classifier} corresponding to digit $d$ (see example use case in Fig.~\ref{fig:auto_exploration}).
This tool can easily be extended to domains other than digits, by using the appropriate classifiers.

Besides classifier-driven exploration, we also borrow objective functions from \cite{bahat2019explorable_sr} and modify them for the JPEG decompression case, as well as add several JPEG-specific objectives to allow tuning local hue and saturation. The full set of available objective functions facilitates a wide range of operations, including manipulating local image variance (\eg using $f(\cdot)=(\text{Var}(\cdot)-c)^2$ for some desired variance level $c$), performing piece-wise smoothing (\eg using $f(\cdot)=\text{TV}(\cdot)$), propagating patches from source to target regions, modifying periodic patterns and more.

Another particularly useful group of tools allows embedding many forms of graphical user input, including various scribbling tools (similar to Microsoft-Paint), modifying local image brightness and even imprinting visual content from an external image. These tools act in two phases (corresponding to the middle pair of images in Fig.~\ref{fig:editing_teaser}). They first enforce consistency of the desired input with the compressed image code, by projecting the scribbled (or imprinted) image onto the set of images that are consistent with the compressed code $x_Q$. Namely, each block of DCT coefficients $X^{\text{scribbled}}_D$ of the scribbled image is modified into
\begin{equation}\label{eq:consistent_scribble}
X^{\text{scribbled}}_D \!\leftarrow\! \left(\text{clip}_{[-\frac{1}{2},\frac{1}{2}]}\!\left(X^{\text{scribbled}}_D\!\oslash\! M\!-\!X_Q\right)\!+\!X_Q\right)\!\odot\! M.
\end{equation} 
This is the left of the middle pair in Fig.~\ref{fig:editing_teaser}. In the second phase, an optimization process over $z$ traverses the learned natural image manifold, searching for the output image that is closest to the consistent scribbled input. This is the right of the middle pair in Fig.~\ref{fig:editing_teaser}. Variants of these tools provide many other features, including automatically searching for the most suitable embedding location, from a consistency standpoint.  Please refer to the supplementary material for detailed descriptions of all tools provided by our GUI.

Our exploration framework is applicable in many domains and use cases, which we demonstrate through a series of representative examples\footnote{Compressed images in our examples are produced by applying the JPEG compression pipeline to uncompressed images, though our method is designed to allow exploration of existing compressed codes.}. Fig.~\ref{fig:dunes} depicts a visually unsatisfying decoded JPEG image (left). Utilizing an artifact removal method yields some improvement, but significant improvement is achieved by allowing a user to edit the image, harnessing specific prior knowledge about the appearance of sand dunes. Another important application involves exploring the range of plausible explanations to the compressed image code, like the different appearances of the shirt in Fig.~\ref{fig:teaser} or the focus of the kitten's attention in Fig.~\ref{fig:kitten_and_fly}. Our framework can also be used to investigate which details could have comprised the original image. This is particularly important in medical and forensic settings. We demonstrate examples of exploring corrupted text in Fig.~\ref{fig:auto_exploration}, and examining a mole in a medical use case in Fig.~\ref{fig:applications}. More examples can be found in the supplementary.
\section{Experiments}\label{sec:experiments}
This work primarily focuses on introducing explorable image decompression, which we demonstrate on various use cases. 
Nevertheless, outputs of our framework are perceptually pleasing even prior to any user exploration or editing, as we show in Fig.~\ref{fig:visual_comparison} in comparison with DnCNN \cite{zhang2017dncnn} and AGARNet \cite{kim2020per_pixel_qf_est}, the only AR methods handling a range of compression levels (like ours) whose code is available online.
Fig.~\ref{fig:quantitative_comparison} (left) further demonstrates the perceptual quality advantage of our method (Ours, GAN), by comparing 
NIQE \cite{mittal2012niqe} perceptual quality scores\footnote{
The no-reference NIQE measure is most suitable for our case, as it does not take into account the GT uncompressed images corresponding to the input JPEG code, which are as valid a decoding as any other output of our network, thanks to its inherent consistency.
Nonetheless, the advantage of our method remains clear even when considering full-reference perceptual quality metrics, as we show in the Supp. using LPIPS \cite{zhang2018lpips}.
}, evaluated on the BSD-100 \cite{Martin2001bsd} dataset.
Finally, our consistency-preserving architecture produces state of the art results even when tackling the traditional (non-diverse, non-explorable) compression artifact removal task.
Fig.~\ref{fig:quantitative_comparison} (right) presents a comparison between
DnCNN and AGARNet, which aim for minimal distortion, and a variant of our model trained using only the $L_1$ penalty, \ie the initialization phase in Sec.~\ref{sec:decoder_design}. 
Our model (Ours, $L_1$) compares favorably to the competition, especially on severely compressed images (low QFs).
Please refer to the Supp. for additional details and comparisons, including evaluation on the LIVE1 \cite{sheikh2005live} dataset.
\section{Conclusion}
We presented a method for user-interactive JPEG decoding, which allows exploring the set of naturally looking images that could have been the source of a compressed JPEG file. Our method makes use of a deep network architecture which operates in the DCT domain, and guarantees consistency with the compressed code by design. A control input signal
enables traversing the set of natural images that are consistent with the compressed code.
Like most decompression works, our framework is tailored to JPEG. However, the proposed concept is general, and can be applied to other compression standards, by identifying and addressing their respective sources of information loss.
%
We demonstrated our approach in various use cases, showing its wide applicability in creativity, forensic, and medical settings.
\vspace{-0.3cm}
\paragraph{Acknowledgements} This research was supported by the Israel Science Foundation (grant 852/17) and by the Technion
Ollendorff Minerva Center.

{\small
\bibliographystyle{unsrt}
\bibliography{Explorable_JPEG}
}
\ifarXiv
\appendixpage
\appendix
\section{Performance Comparison}\label{sec:performance}
\begin{figure*}[t]
\centering
\includegraphics[width=\columnwidth]{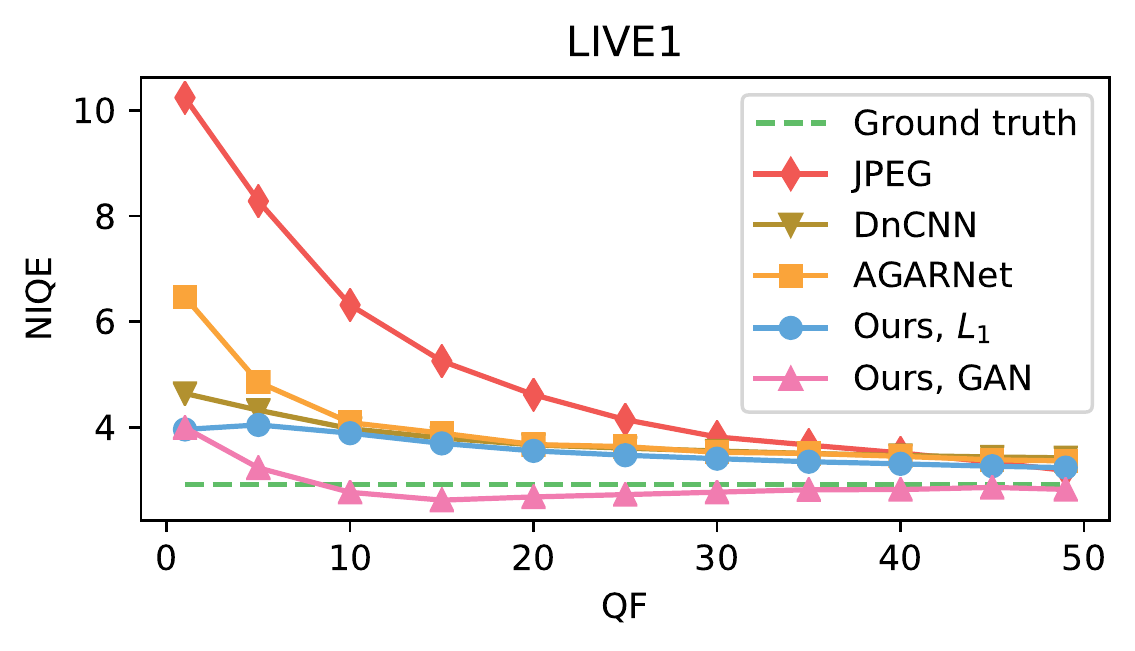}%
\includegraphics[width=\columnwidth]{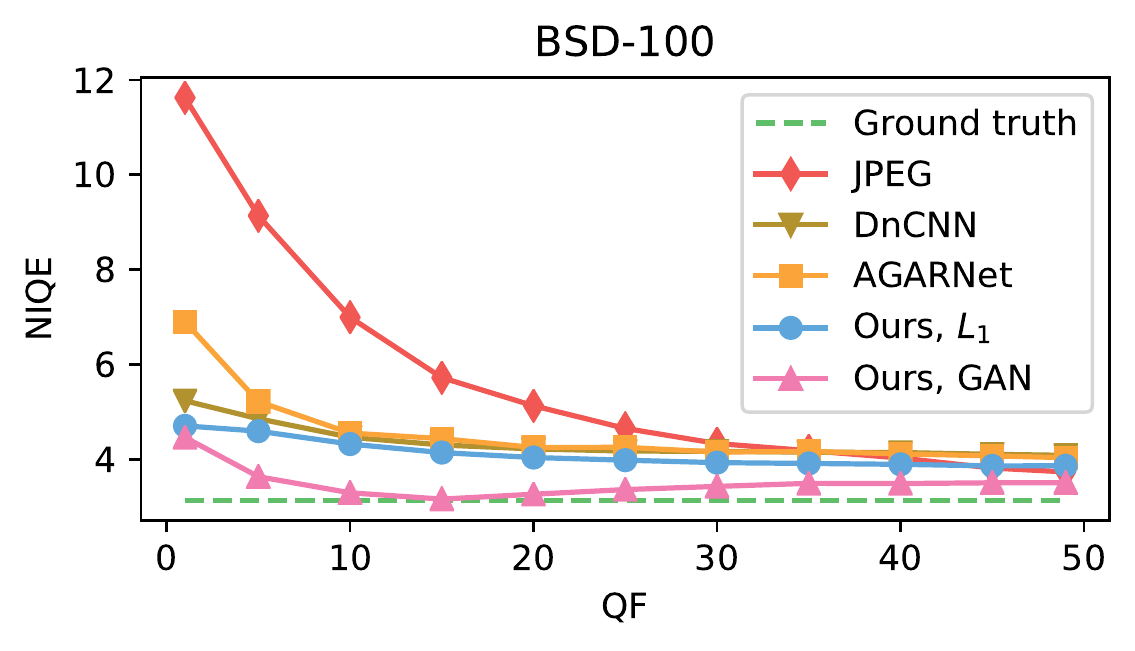}
{Perceptual quality (no-reference, lower is better)}
\centering
\includegraphics[width=\columnwidth]{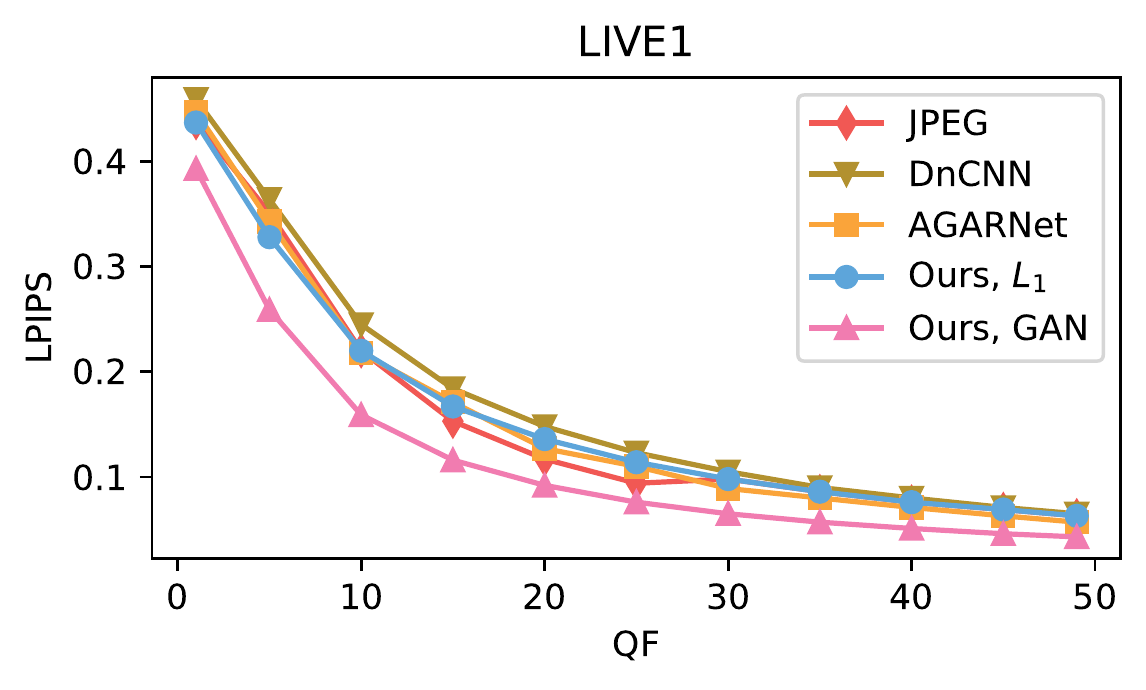}%
\includegraphics[width=\columnwidth]{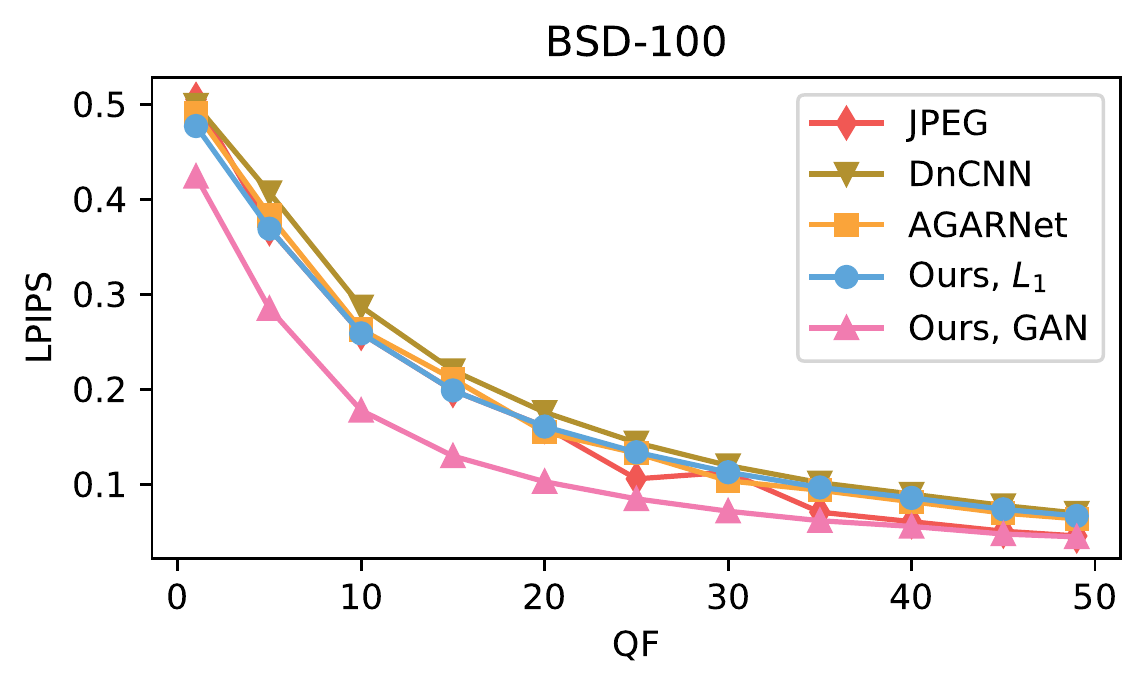}
{Perceptual quality (full-reference, lower is better)}
\centering
\includegraphics[width=\columnwidth]{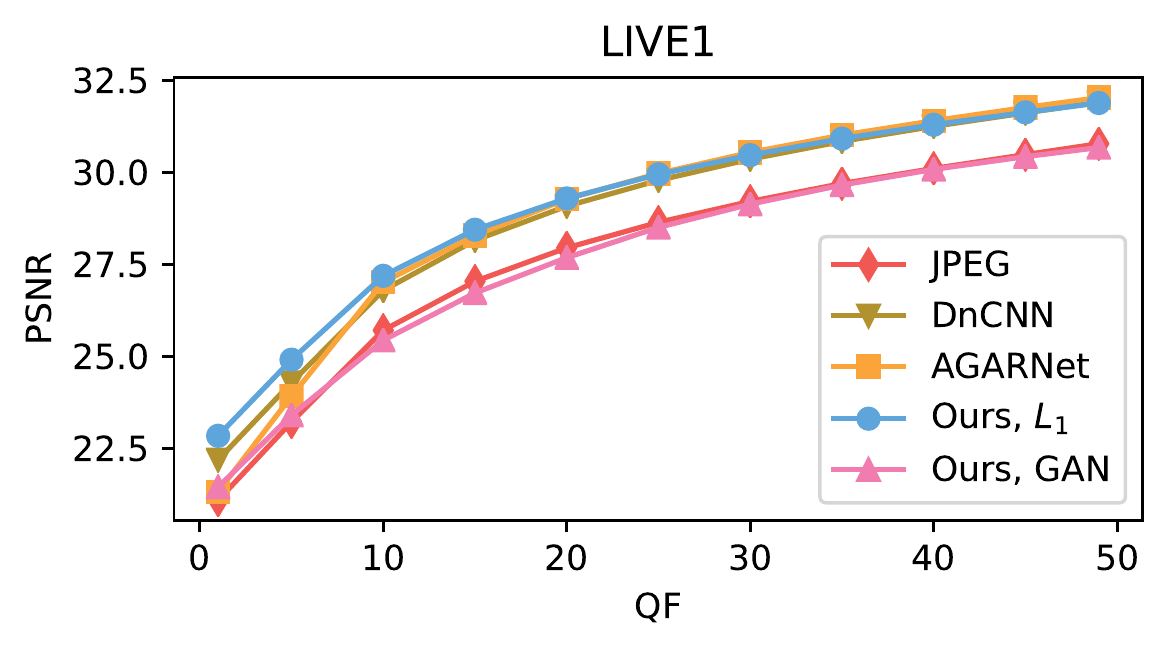}%
\includegraphics[width=\columnwidth]{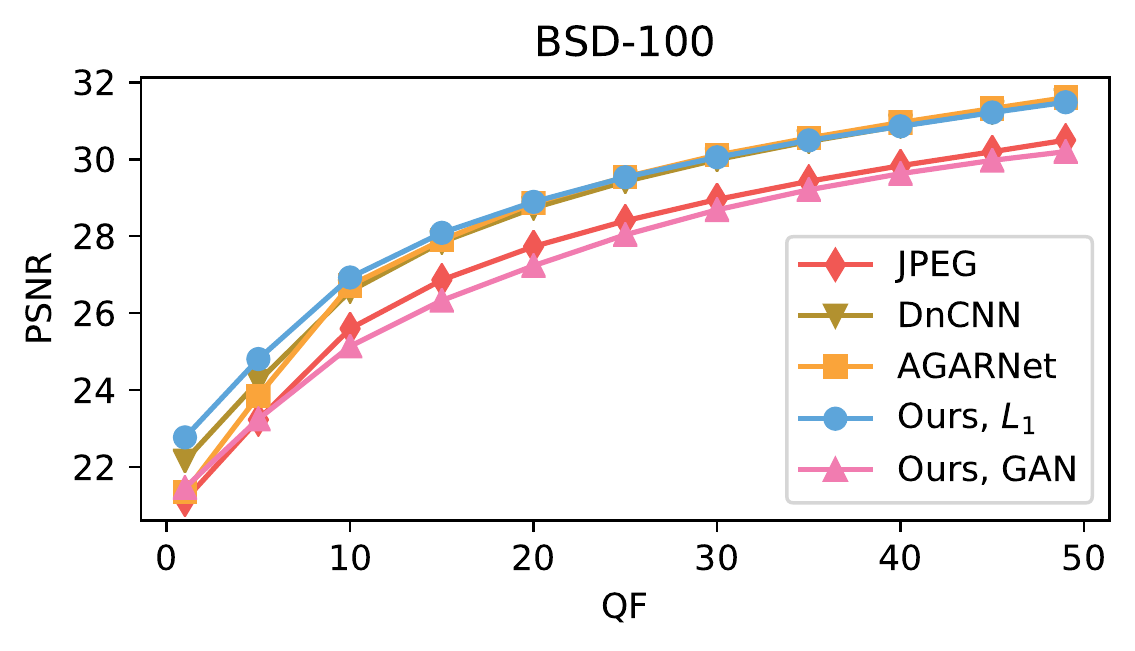}
{Reconstruction quality (full-reference, higher is better)}
\caption{\label{fig:quantitative_comparison_supp_color}\textbf{Quantitative performance evaluation - color images.} Comparing performance on the LIVE1 \cite{sheikh2005live} (left column) and BSD-100 \cite{Martin2001bsd} (right column) datasets, in terms of no-reference perceptual quality (top row) and full reference perceptual quality (middle row) and image distortion (bottom row), using the NIQE ($\downarrow$), LPIPS ($\downarrow$) and PSNR ($\uparrow$) metrics, respectively. Please see details in Sec.~\ref{sec:performance}.}
\end{figure*}
\begin{figure*}[t]
\centering
\includegraphics[width=\columnwidth]{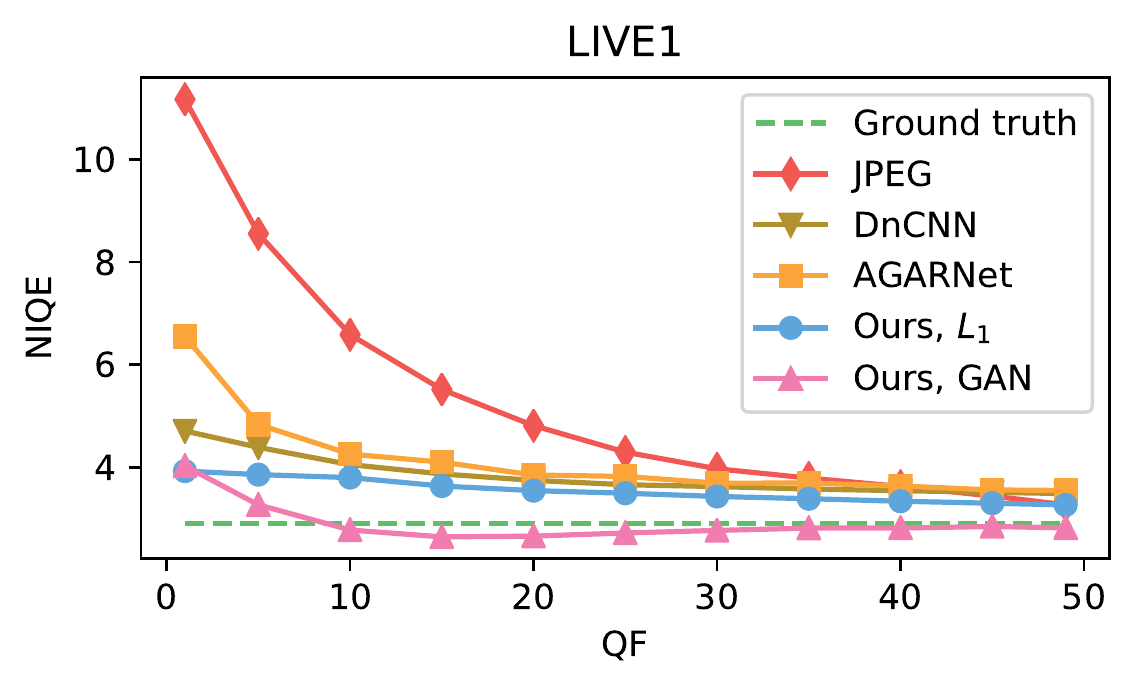}%
\includegraphics[width=\columnwidth]{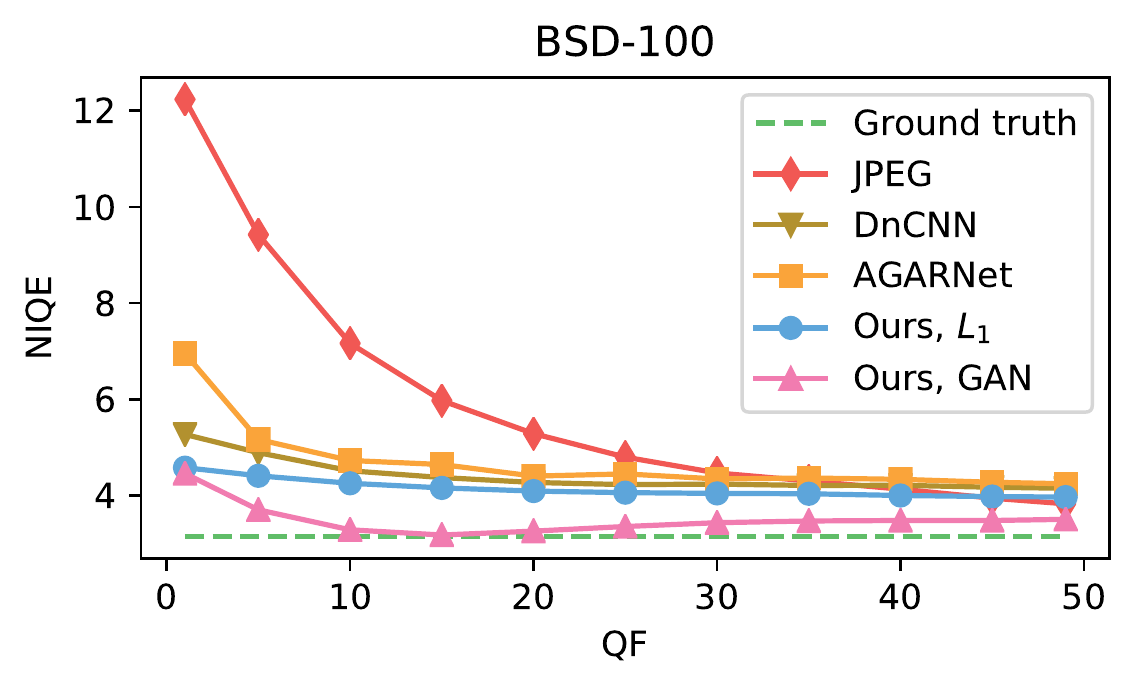}
{Perceptual quality (no-reference, lower is better)}
\centering
\includegraphics[width=\columnwidth]{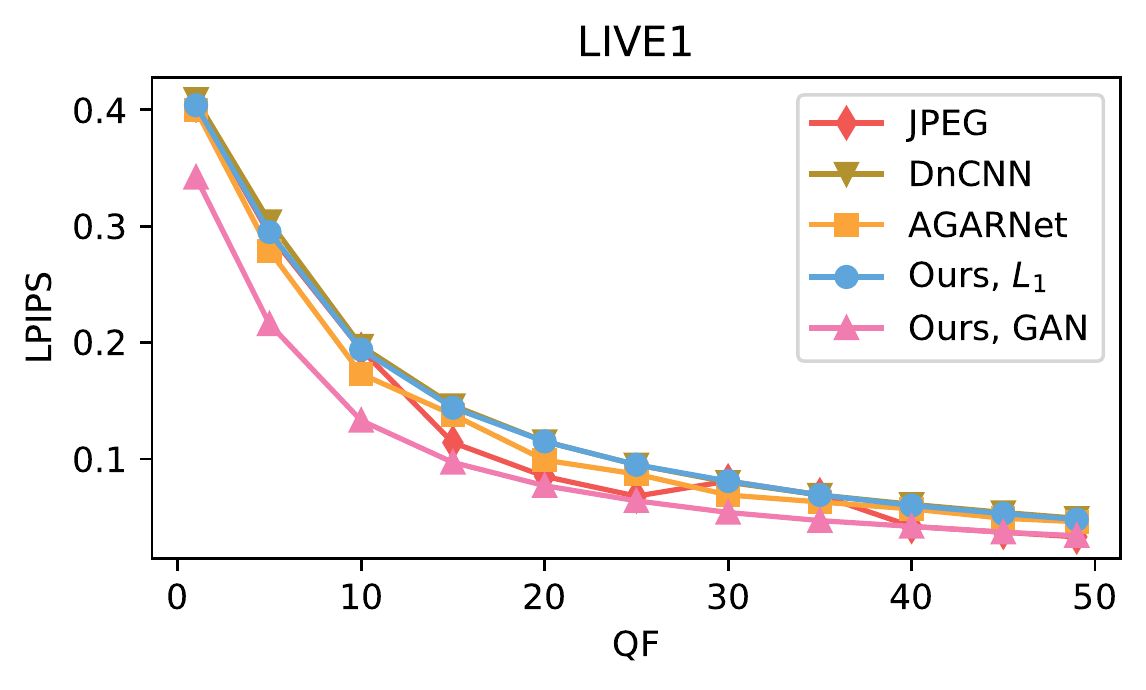}%
\includegraphics[width=\columnwidth]{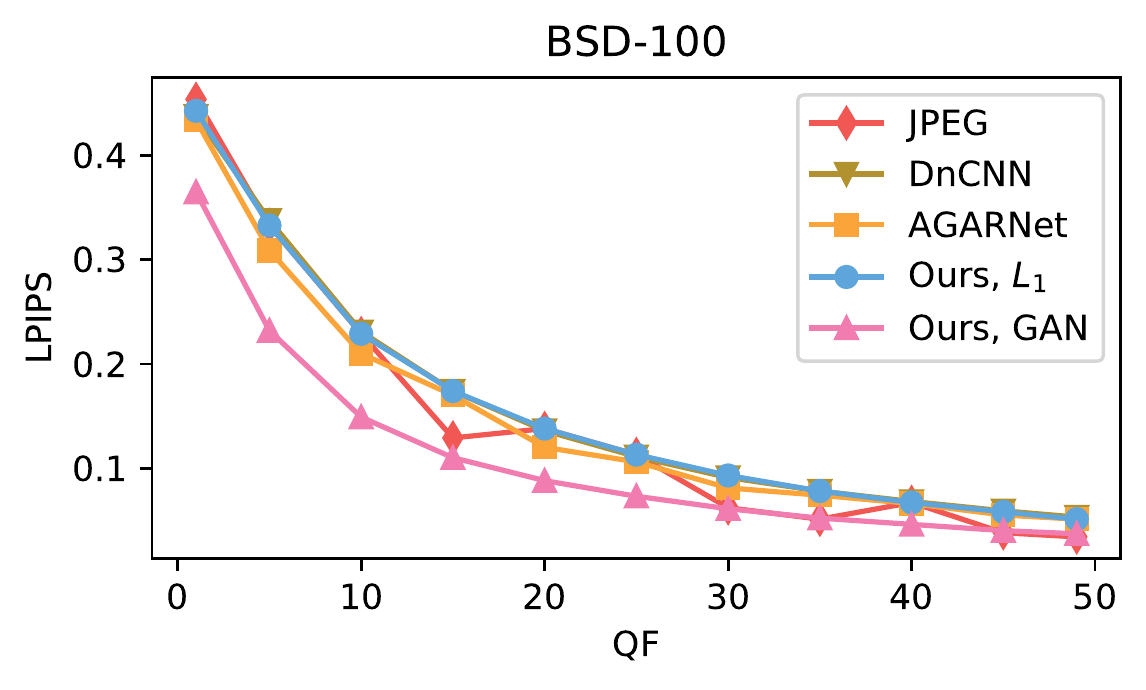}
{Perceptual quality (full-reference, lower is better)}
\centering
\includegraphics[width=\columnwidth]{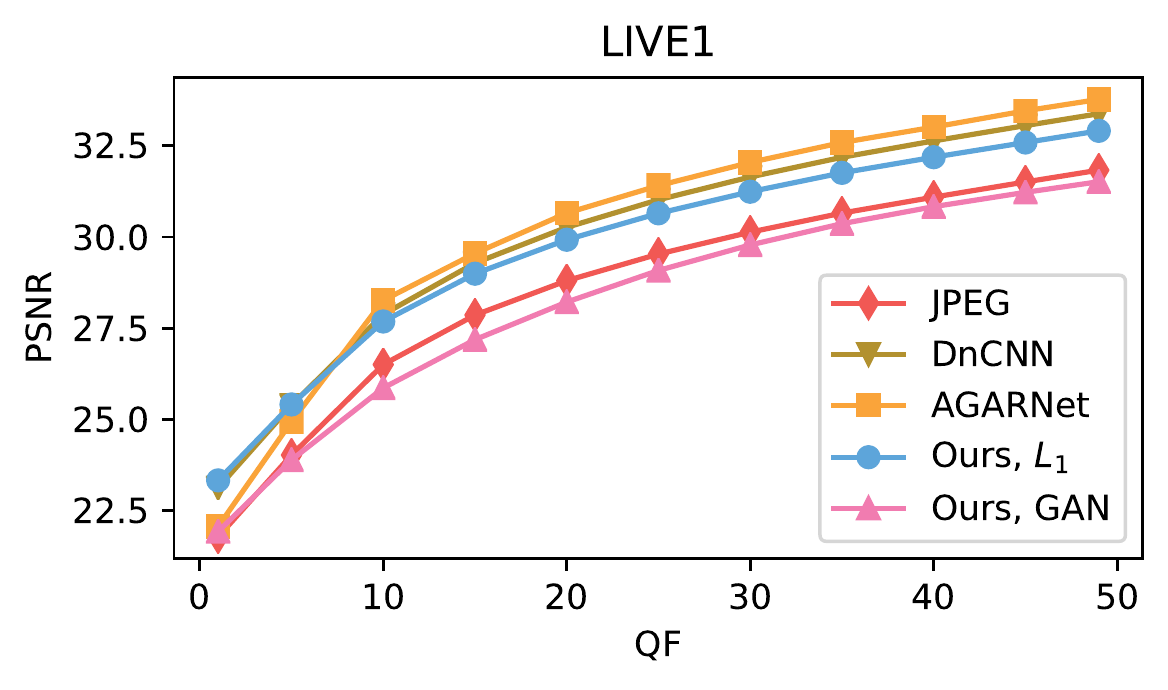}%
\includegraphics[width=\columnwidth]{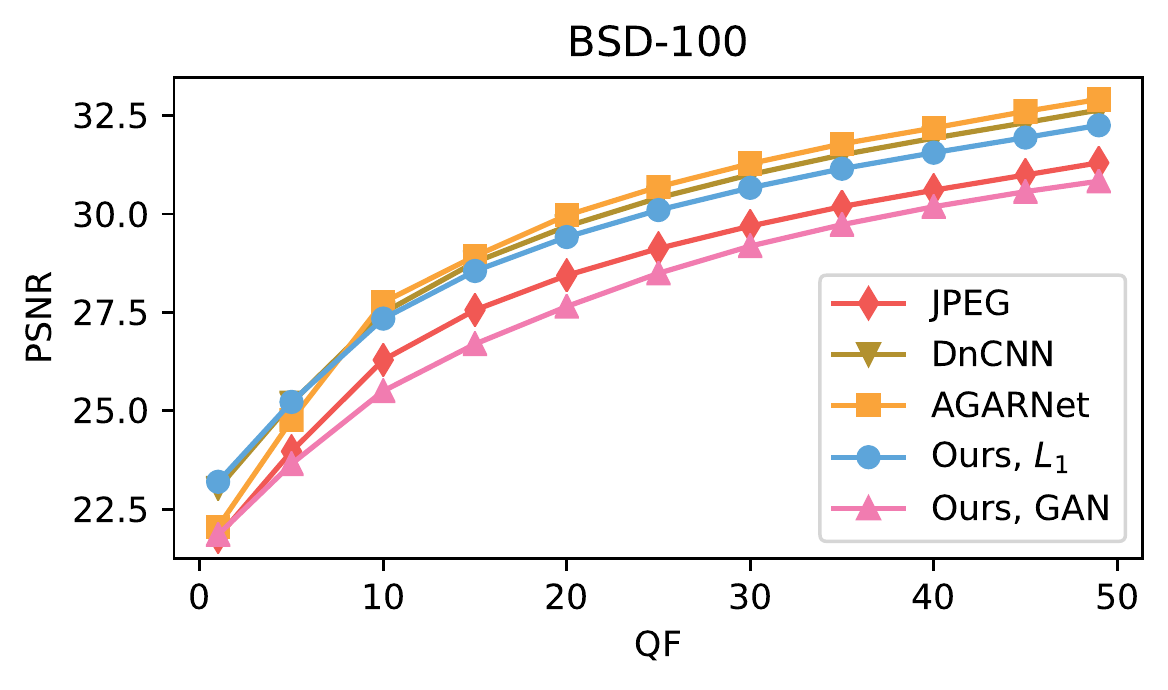}
{Reconstruction quality (full-reference, higher is better)}
\caption{\label{fig:quantitative_comparison_supp_grayscale}\textbf{Quantitative performance evaluation - grayscale images.} 
Comparing performance on the LIVE1 \cite{sheikh2005live} (left column) and BSD-100 \cite{Martin2001bsd} (right column) datasets, in terms of no-reference perceptual quality (top row) and full reference perceptual quality (middle row) and image distortion (bottom row), using the NIQE ($\downarrow$), LPIPS ($\downarrow$) and PSNR ($\uparrow$) metrics, respectively. Please see details in Sec.~\ref{sec:performance}.
}
\end{figure*}
Our JPEG decoding framework is the first to facilitate exploration of the abundant plausible images corresponding to a given compressed JPEG code, and therefore cannot be compared to any existing method. Nonetheless, it produces high quality outputs even prior to applying any user editing. To evaluate the quality of pre-edited outputs and compare it with that of existing JPEG artifact removal methods, we perform experiments using two datasets commonly used for evaluating artifact removal, namely the LIVE1 \cite{sheikh2005live} and BSD-100 \cite{Martin2001bsd} datasets, containing 29 and 100 images, respectively.

Methods for removing JPEG artifacts strive to achieve one of two possible goals; either they attempt to minimize outputs' distortion with respect to corresponding ground truth (GT) uncompressed images, or they try to maximize their outputs' perceptual quality. We quantitatively evaluate the performance with respect to each of these different goals, by adapting the commonly used metric for each goal: Distortion is evaluated by measuring peak signal to noise ratio (PSNR) w.r.t. GT images, while perceptual quality is evaluated using both the naturalness image quality evaluator (NIQE, \cite{mittal2012niqe}) no-reference score, and the learned perceptual image patch similarity (LPIPS, \cite{zhang2018lpips}) full-reference score.
We compare our method with DnCNN \cite{zhang2017dncnn} and AGARNet \cite{kim2020per_pixel_qf_est}, the only AR methods handling a range of compression levels (like ours) whose code is available online. As these methods aim for minimal distortion rather than perceptual quality, we train a variant of our model using the $L_1$ penalty instead of the full penalty in~\eqref{eq:total_loss}, which is based on adversarial loss.
Our $Y$ channel reconstruction $L_1$ model uses $370$ output channels for each convolution operation but the last, instead of the $320$ channels used for our perceptually oriented $Y$ channel model. Like AGARNet, we train this variant using a wider range of QFs, spanning from $1$ to $90$. To allow color image evaluation, we couple our $L_1$ model with an $L_1$-trained variant of our chroma reconstruction model as well.
We consider our $L_1$-minimized model and our main model as two different configurations, denoting them by ``Ours, $L_1$'' and ``Ours, GAN'', respectively. 
Since the available pretrained AGARNet model only handles single channel images (only the $Y$ channel), for evaluating color images we augment their reconstructed $Y$ channel with the $C_b$ and $C_r$ channels decoded by the standard JPEG pipeline.

Note that our main (GAN) model takes an additional input signal $z$, facilitating user exploration. However, evaluating post-exploration outputs is an ill-defined task that mostly depends on the user's intentions. We circumvent this problem by drawing $50$ random $z$ signals per image, then averaging PSNR, NIQE and LPIPS scores over the resulting $50$ outputs per image, over the entire evaluation set.

Quantitative evaluations of both metrics on color and grayscale images, on both datasets, are presented in Figs.~\ref{fig:quantitative_comparison_supp_color} and~\ref{fig:quantitative_comparison_supp_grayscale} respectively.
We present scores corresponding to the standard JPEG decompression in all our comparisons, and for the no-reference NIQE scores present also the values corresponding to the GT images.
The results on color images (Fig.~\ref{fig:quantitative_comparison_supp_color}) indicate that our distortion minimizing model (blue), trained to minimize distortion, compares favorably both with DnCNN (brown) and with AGARNet (orange) in terms of reconstruction error (bottom row), on both datasets, especially on severely compressed images (low QFs). When evaluating on grayscale images (Fig.~\ref{fig:quantitative_comparison_supp_grayscale}), this model is on par with the competition. As can be expected, PSNR scores of our main model (``Ours, GAN''), trained for perceptual quality, are significantly lower, even slightly below JPEG image scores.

As for perceptual quality, NIQE scores (where lower is better) on the top rows suggest that our GAN-trained model (pink) performs well across all evaluated QFs and both datasets, 
obtaining similar scores to those of the GT images. Evaluating using the full-reference LPIPS score (middle rows) also indicates high perceptual quality.
As expected, both competing methods and our distortion minimizing model perform significantly worse, as they were all trained to minimize distortion.

Finally, the advantage of our method in terms of perceptual quality is evident in a qualitative (visual) comparison, as presented in Fig.~\ref{fig:visual_comparison} for the case of severely compressed images (QF=$5$). This advantage in perceptual quality persist across different compression levels, as demonstrated using a more moderate compression level (QF=$10$) in Fig.~\ref{fig:visual_comparison_QF10}.
We do not present the corresponding uncompressed GT images in these figures, since the GT images are as valid a decoding as any other output of our network, due to its inherent consistency with the input code. For the curious readers however, we present these corresponding GT images in Fig.~\ref{fig:visual_comparison_GT}.
\begin{figure*}[th]
\centering
\includegraphics[width=\textwidth]{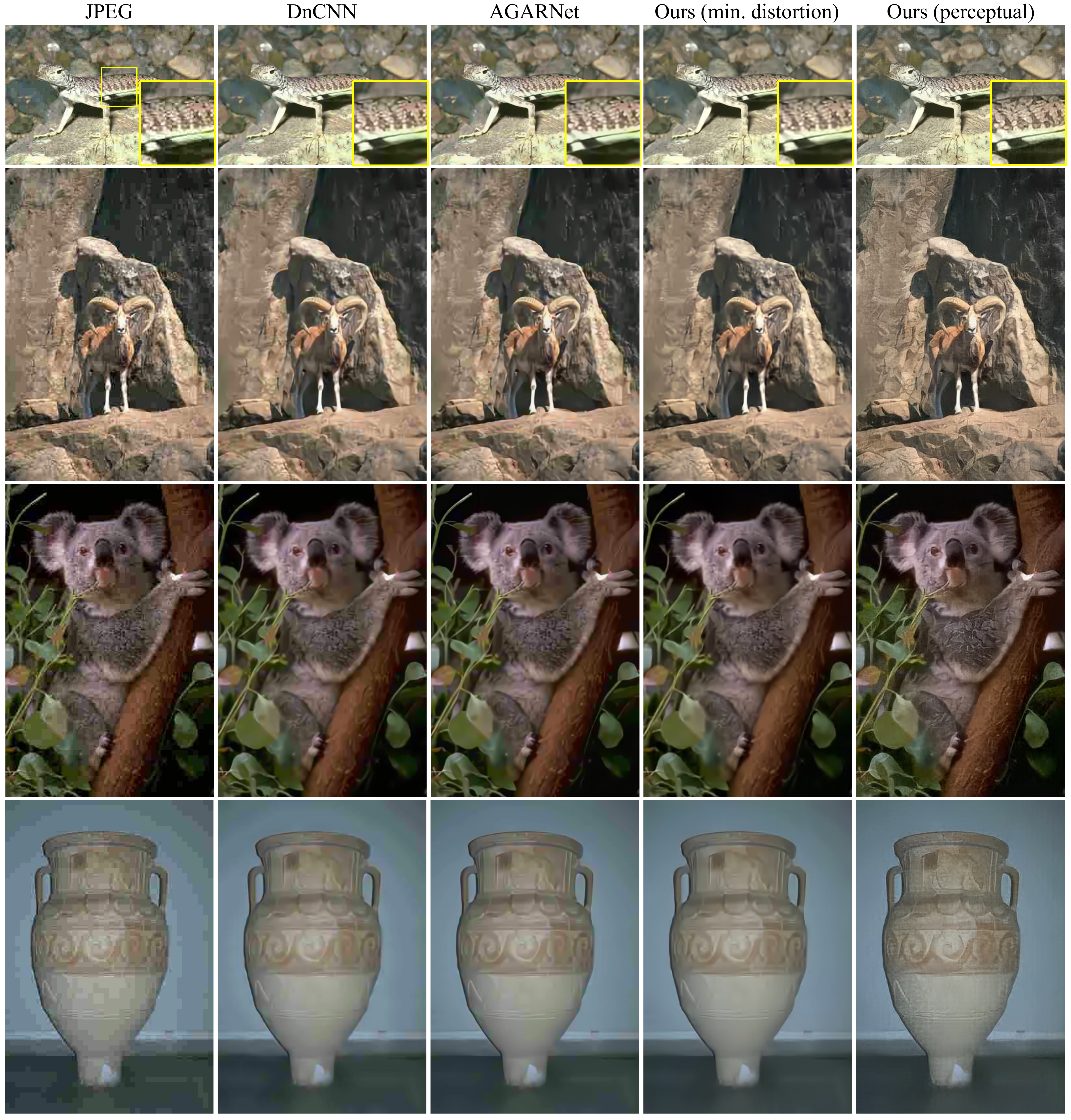}
\vspace{-10pt}\caption{\label{fig:visual_comparison_QF10}\textbf{Qualitative comparison using QF=$\textbf{10}$.} 
From left to right: Outputs by the JPEG decompression pipeline, DnCNN \cite{zhang2017dncnn} AR method, AGARNet \cite{kim2020per_pixel_qf_est} AR method, a variant of our model trained to minimize distortion and our GAN-trained model.
Similar to the results presented in Fig.~\ref{fig:visual_comparison} for the harsher compression level (QF=$5$), our GAN-trained model (right) produces sharper and more photo-realistic outputs for the QF=$10$ case as well, supporting quantitative findings in Figs.~\ref{fig:quantitative_comparison_supp_color} and~\ref{fig:quantitative_comparison_supp_grayscale}, which indicate a significant perceptual quality advantage across a range of compression levels. Images taken from the BSD-100 \cite{Martin2001bsd} test set.
}
\end{figure*}
\begin{figure*}[th]
\centering
\includegraphics[width=\textwidth]{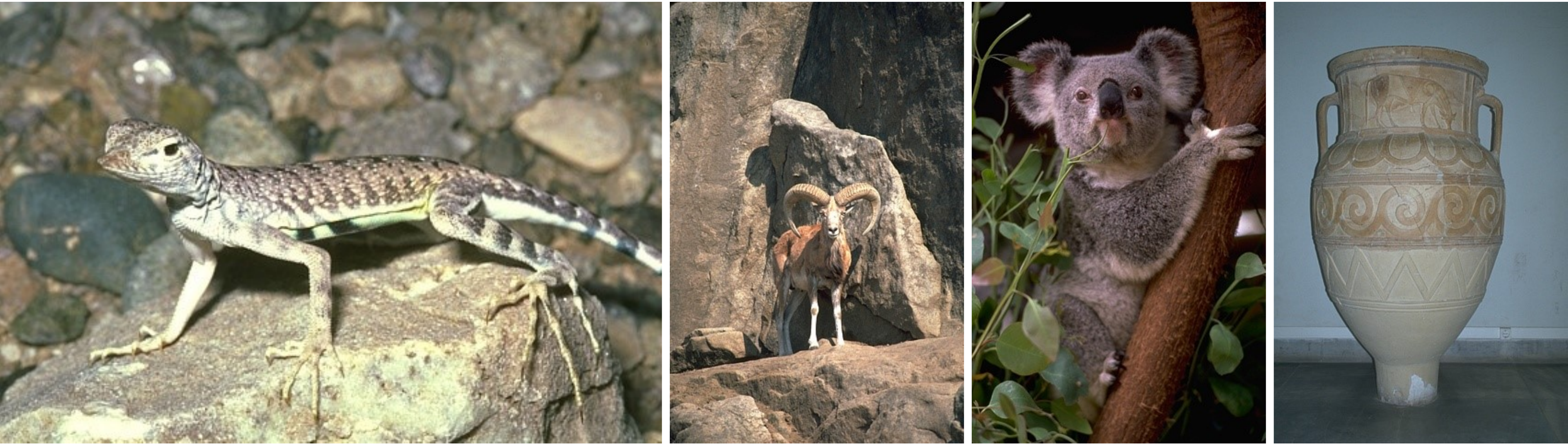}
\vspace{-10pt}\caption{\label{fig:visual_comparison_GT}\textbf{Ground truth uncompressed images corresponding to images in Figs.~\ref{fig:visual_comparison} and~\ref{fig:visual_comparison_QF10}.} 
}
\end{figure*}
\section{Exploration Tools}
\begin{figure}[b]
\centering
\includegraphics[width=\columnwidth]{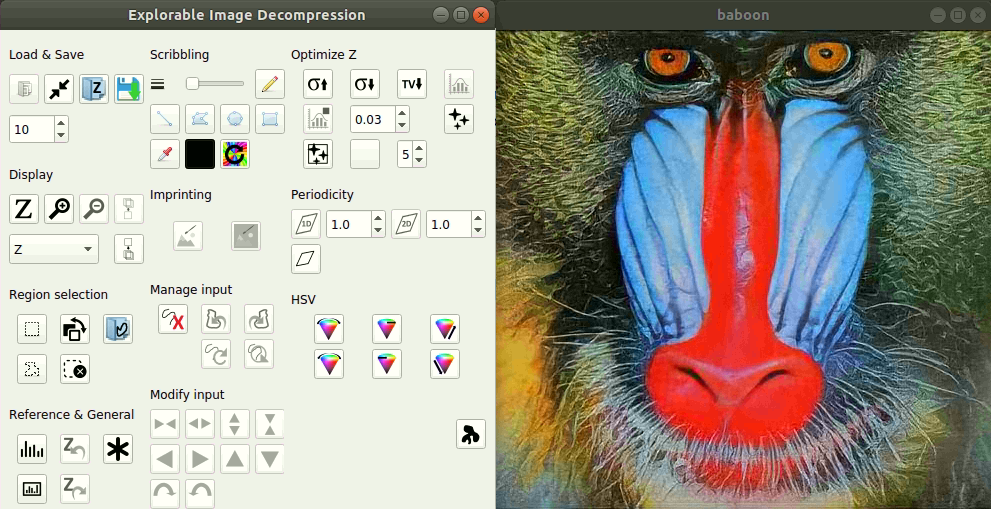}
\vspace{-10pt}\caption{\label{fig:GUI}\textbf{Our Exploration GUI.} Performing explorable image decompression of a compressed Mandrill image, using our graphical user interface.}
\end{figure}
Our framework's GUI (depicted in Fig.~\ref{fig:GUI}) comprises many editing and exploration tools that facilitate intricate editing operations. As we explain in Sec.~\ref{sec:editing_tools}, these tools work by triggering an optimization process over the space of control signals $z$, optimizing one of several possible objective functions $f(\cdot)$.
This is analogous to traversing the manifold of perceptually plausible images learned by our network, while always remaining consistent with the compressed image code. 
We introduce a novel automatic exploration tool, allowing users to explore solutions of interest at a press of a button, as well as some JPEG-specific tools. In addition, we incorporate most editing tools introduced for explorable super resolution \cite{bahat2019explorable_sr}, by adapting them to the JPEG decompression case.

Editing can be applied to the entire image, or to a specific region marked by the user. Some tools enable more precise editing, by employing Microsoft-Paint-like buttons, including pen and straight line (with adjustable line width), as well as polygon, square and circle drawing tools. 

We denote an output image prior to minimizing each objective $f$ by $\hat{x}_0=\psi(x_Q,z_0)$, where signal $z_0$ is either a neutral (pre-editing) control signal or the result of a prior editing process. Note that any function $f(\cdot)$ computed on the entire image can alternatively be computed on a specific region thereof, by masking out the rest of the image. We use $P(\cdot)$ to denote a patch extraction operator\footnote{Objective functions operating on image patches (rather than directly on the image itself) use partially overlapping $6\times6$ patches. The degree of overlap varies, and indicated separately for each tool.}, for those objective functions below that expect this kind of input. We next describe the different available objective functions and the way they are utilized in our GUI. 


\subsection{Variance Manipulation}\label{sec:var_manipulation}
This is a set of tools which operates by manipulating the local variance of all partially overlapping image patches in the selected region. We employ cost function \mbox{$f(\hat{x})=\sum(\text{Var}(P(\hat{x}))-\text{Var}(P(\hat{x}_0))\pm\delta)^2$},  where the sum runs over all overlapping patches, and optimize over $z$ to modify (increase or decrease) the local, per-patch variance by a desired value $\delta$. The effect of applying this tool is demonstrated in Fig.~\ref{fig:tool_demos}(a) between the left and middle images.
\subsection{Encouraging Piece-wise Smoothness}
This tool acts by minimizing the total variations (TV) in an image or a region: \mbox{$f(\hat{x})=\text{TV}(\hat{x})$}. In particular, we minimize the sum of absolute differences between each pixel in the image and its $8$ neighboring pixels. This function can be minimized for a single region, or simultaneously minimized for several marked image areas. The effect of applying this tool is demonstrated in Fig.~\ref{fig:tool_demos}(a) between the middle and right images.

\subsection{Imposing Graphical User Input}\label{sec:graphical_input}
Our GUI comprises a large set of tools to allow imposing a graphical user input on the output image, by minimizing \mbox{$f(\hat{x})=\|\hat{x}-x^{\text{scribbled}}\|_1$}. 
The desired graphical content $x^{\text{scribbled}}$ is imposed in a consistency preserving manner, by projecting it onto the set of images that are consistent with the compressed code $x_Q$. Namely, each block of DCT coefficients $X^{\text{scribbled}}_D$ of the desired input is modified by applying Eq.~(\ref{eq:consistent_scribble}), repeated here for fluency:
\begin{equation}
X^{\text{scribbled}}_D \leftarrow \left(\text{clip}_{[-\frac{1}{2},\frac{1}{2}]}\left(X^{\text{scribbled}}_D\oslash M-X_Q\right)+X_Q\right)\odot M.
\end{equation}
The modified $X^{\text{scribbled}}$ (depicted, \emph{e.g.}, in the left of the middle pair of images in Fig.~\ref{fig:editing_teaser}) is already consistent with the compressed input code $x_Q$. The user then has the option of translating, resizing or rotating the inserted content using arrow buttons, while consistency is re-enforced automatically after each of these operations. Editing is completed when the user initiates the optimization process, traversing the $z$ space looking for the image $\hat{x}$ that is closest to the desired consistent content $X^{\text{scribbled}}$, while lying on the learned manifold of perceptually plausible images.

The desired input $x^{\text{scribbled}}$ can originate from any of the following sources:
\begin{enumerate}
    \item \emph{User scribble:} A user can use the Microsoft-Paint-like drawing tools, where scribbling color can be chosen manually or sampled from any given image (including the edited one). Please see Fig.~\ref{fig:editing_mole} for an example usage of this tool.
    \item \emph{Manipulating HSV:} Local hue, saturation and relative brightness (value) of $\hat{x}$ can be manipulated by using one of $6$ designated buttons. This results in a desired appearance $x^{\text{scribbled}}$, whose consistency is continuously enforced after each button press, by computing \eqref{eq:consistent_scribble}. Brightness manipulation was already facilitated in \cite{bahat2019explorable_sr} for small image details, but larger regions could not be manipulated, as their HSV attributes are strictly determined by the low-resolution input. In contrast, JPEG compression often discards information corresponding to these attributes, thus allowing and necessitating their exploration. 
    \item \emph{Imprinting:} A user can import graphical content, either from within the edited image or from an external one, and then enforce it on $\hat{x}$. The user first selects the desired content to import, and then marks the target region's bounding rectangle on $\hat{x}$. JPEG compression operates on $8\times 8$ pixel blocks, making it highly sensitive to small image shifts. Therefore, to adapt this tool from \cite{bahat2019explorable_sr}, we propose an option to automatically find a sub-block shifting of the imported content, that yields the most consistent imprinting. Please see Fig.~\ref{fig:editing_teaser} for an example usage of this tool.
\end{enumerate}
\paragraph{Subtle region shifting}
A variant of the imprinting tool allows applying subtle local affine transformations. It works by imprinting the region of interest onto itself, then allowing a user to utilize the shifting, resizing and rotating buttons to modify the selected region from its original appearance, before triggering the final $z$ optimization process.

\begin{figure}[t]
\centering
\includegraphics[width=\columnwidth]{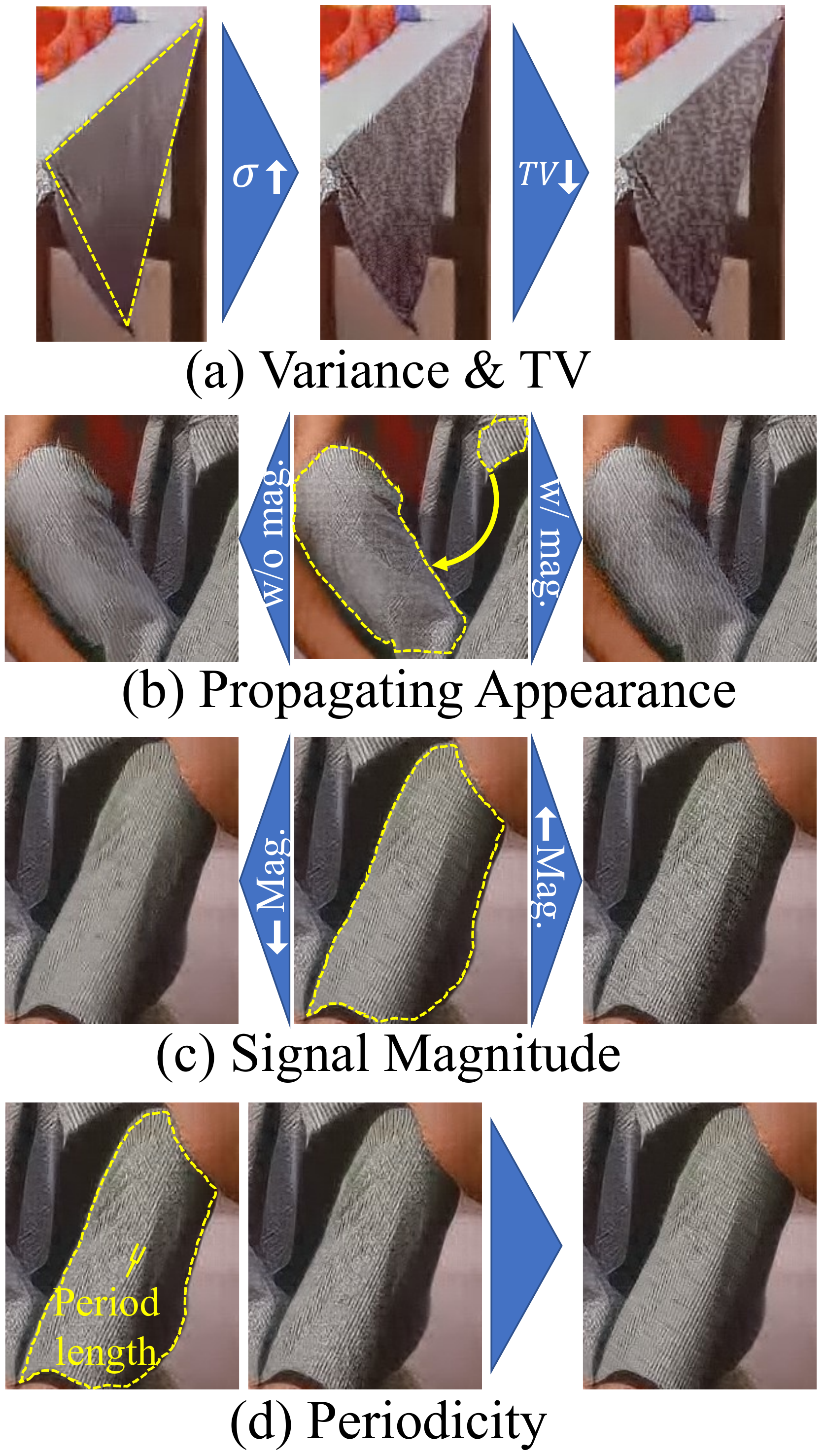}
\vspace{-10pt}\caption{\label{fig:tool_demos}\textbf{Exploration tool examples.} (a) A pre-edited region (left) is manipulated to have higher local variance (middle). Then TV minimization is applied for a smoother result (right). (b) Source and target regions are marked by the user (middle). Appearance of patches is then propagated from source to target, while keeping (right) or ignoring (left) source patches signal magnitude. (c) A desired region (middle) is manipulated by increasing (right) or decreasing (left) local signal magnitude. (d) Desired region and period length are marked by user (left). The effect of periodicity encouraging tool is visible between middle and right images.  }
\end{figure}

\subsection{Desired Dictionary of Patches}
This tool manipulates target patches in a desired region to resemble the patches comprising a desired source region, either taken from an external image or from a different location in the edited image. The latter case is demonstrated in Fig.~\ref{fig:tool_demos}(b): patches from a small yellow region in the middle image are propagated to the larger yellow region, resulting in the right image.

The corresponding cost function penalizes for the distance between each patch in the target region and its nearest neighbor in the source region.
To allow encouraging desired textures across regions with different colors, we first remove mean patch values from each patch, in both source and target patches. To reduce computational load, we discard some of the overlapping patches, by using $2$ and $4$ rows strides in the source and target regions, respectively. 
This tool was used for creating the result in Fig.~\ref{fig:dunes} (right image), by propagating patches depicting sand-waves from the center of the image to its upper-left regions.

\paragraph{Ignoring patches' variance}
A variant of this tool (demonstrated in left image in Fig.~\ref{fig:tool_demos}(b)) allows encouraging desired textures without changing current local variance. To this end, we normalize patches' variance, in addition to removing their mean. Then while optimizing over $z$, we add an additional penalty that preserves the original variance of each target patch, while encouraging its (normalized) signal to resemble that of its closest (normalized) source patch.

\subsection{Signal magnitude manipulation}
An additional tool operating on image patches attempts to amplify or attenuate the magnitude of the signal in existing patches, while preserving existing patch structures. The effect of this tool is demonstrated in Fig.~\ref{fig:tool_demos}(c), both for increasing (right) and for decreasing (left) the signal magnitude within the region marked on the middle image. Similar to the variance manipulation tool described in Sec.~\ref{sec:var_manipulation}, we use \mbox{$f(\hat{x})=\sum\|\tilde{P}(\hat{x})-(1\pm\delta)\tilde{P}(\hat{x}_0)\|^2$} as our cost function, where the sum runs over all overlapping patches. It penalizes for the difference between the newly constructed image patches and the $(1\pm\delta)$ times magnified/attenuated versions of the corresponding existing patches, where operator $\tilde{P}(\cdot)$ extracts image patches and subtracts their respective mean values.
This tool was also utilized for creating the result in Fig.~\ref{fig:dunes} (right image), by enhancing the sand-wave appearance of patches propagated to the upper left image regions.
\subsection{Encouraging Periodicity}
This tool encourages the periodic nature of an image region, across one or two directions determined by a user. The desired period length (in pixels) for each direction can be manually set by the user, or it can be automatically set to the most prominent period length, by calculating local image self-correlation. Periodicity is then encouraged by penalizing for the difference between the image region and its version translated by a single period length, for each desired direction. 
We used this tool too when creating Fig.~\ref{fig:dunes} (right image), for encouraging the sand-waves to have an approximately constant period length (in the appropriate direction), thus yielding a more realistic appearance. The effect of this tool is also demonstrated in Fig.~\ref{fig:tool_demos}(d), where a desired period length and direction are marked by a user on the left image (yellow curly bracket), as well a region to be manipulated (yellow dashed line). The result after Optimizing over $z$ (right) is a sharper and cleaner appearance of the stripes on the manipulated image region.
\subsection{Random Diverse Alternatives}
This tool allows exploring the image manifold in a random manner, producing $N$ different outputs by maximizing the $L_1$ distance between them in pixel space. These images (or sub-regions thereof) can then serve as a baseline for further editing and exploration.
\paragraph{Constraining distance to current image}
A variant of this tool adds the requirement that all $N$ images should be close to the current $\hat{x}_0$ (in terms of $L_1$ distance in pixel space).
\subsection{Automatic Exploration of Classes}
This novel tool allows exploring a predefined set of solutions of interest to a user-marked region at a press of a button. We exemplify it using the case of the $10$ possible numerical digits $d\in\{0-9\}$. To this end, we use the $d^{\text{th}}$ output of a pre-trained digit classifier as our objective function, $f(\cdot)=\text{Classifier}_d(\cdot)$, and produce $10$ different outputs corresponding to the $10$ digits, by repeatedly maximizing $f$ over $z$ using a different $d\in\{0-9\}$ each time. The obtained digit-wise optimized outputs are then presented to the user, who can, \eg, examine their plausibility to assess the likelihood of the underlying visual content corresponding to each of these digits, as we show in Fig.~\ref{fig:auto_exploration} for the hour digit on the captured smartphone's screen.
This tool can accommodate any type of classifier, and can therefore be very useful in forensic and medical applications (\eg for predicting malignant/benign tumors).
\section{Editing Processes and Additional Examples}
\begin{figure}[t]
\centering
\includegraphics[width=\columnwidth]
{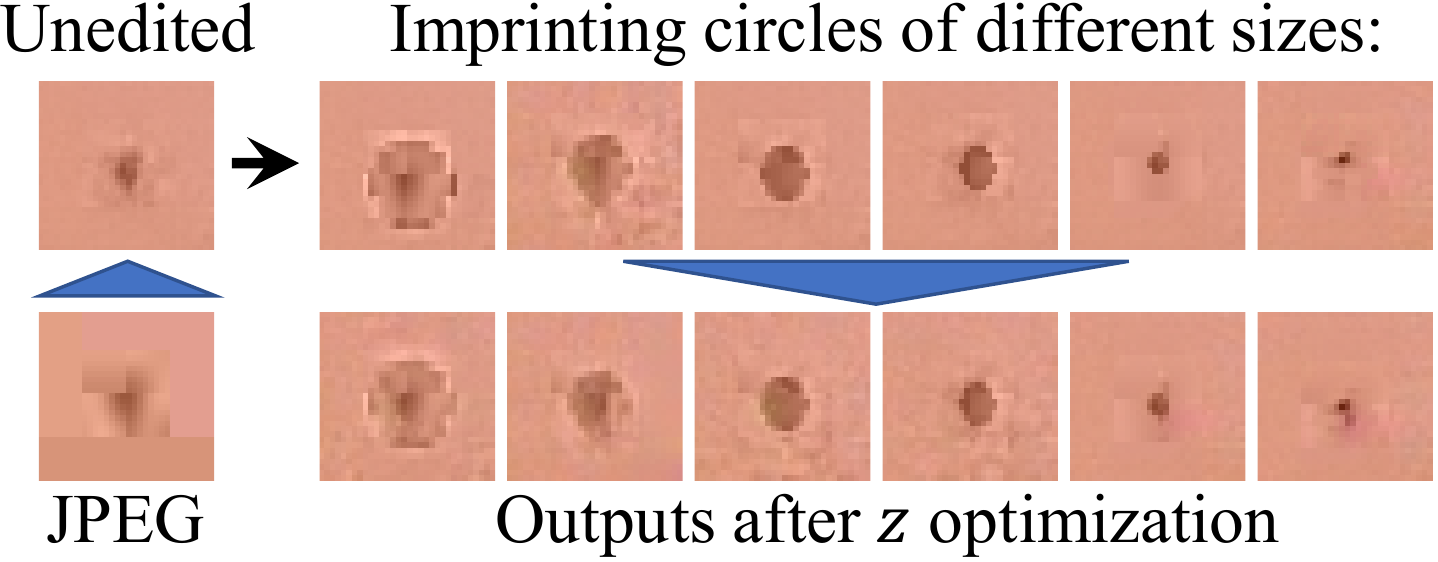}
\caption{\label{fig:editing_mole}\textbf{The making of Fig.~\ref{fig:applications}.}
Here, we imprint brown disks of varying radii on the region of the mole. The top row shows the projection of the naively imprinted image onto the set of consistent solutions. The bottom row shows the final output of our method, after determining the control signal $z$ that causes the net's output to resemble the most to the images in the top row.}
\end{figure}
\begin{figure}[t]
\centering
\includegraphics[width=\columnwidth]{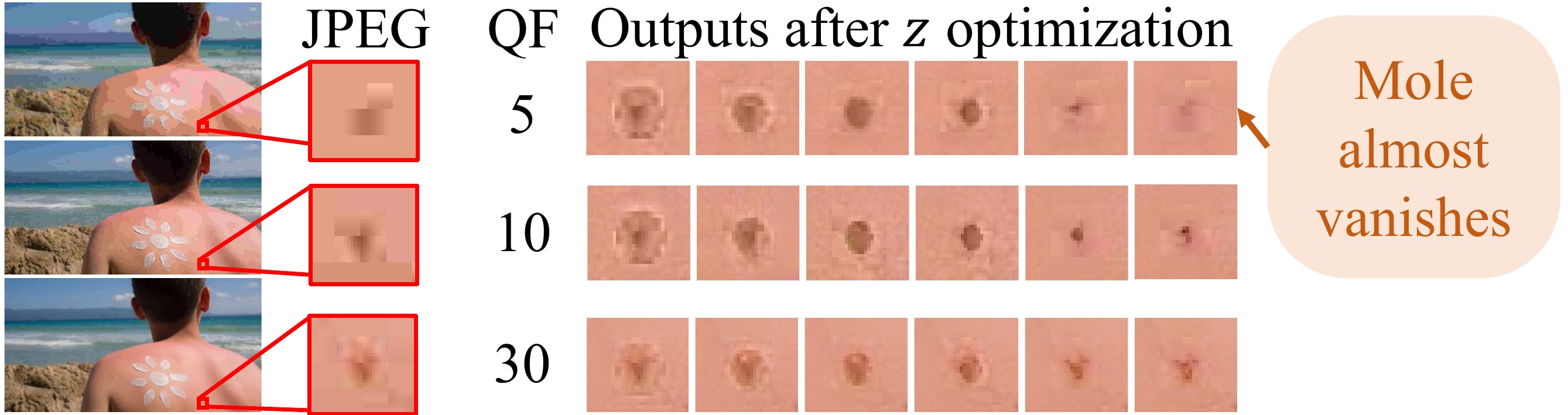}
\caption{\label{fig:different_QFs}\textbf{Effect of QF on exploration space.} The ability to imprint brown disks of different radii on the mole, strongly depends on the QF. When the QF is small (agressive compression), the space of consistent solutions is large. In this case, we can imprint a large mole in one extreme, or completely remove the mole in the other extreme. However, as the QF increases, the space of consistent solutions becomes smaller, and the range of mole sizes that can be imprinted reduces accordingly.}
\end{figure}
\begin{figure*}[ht]
\centering
\includegraphics[width=\textwidth]{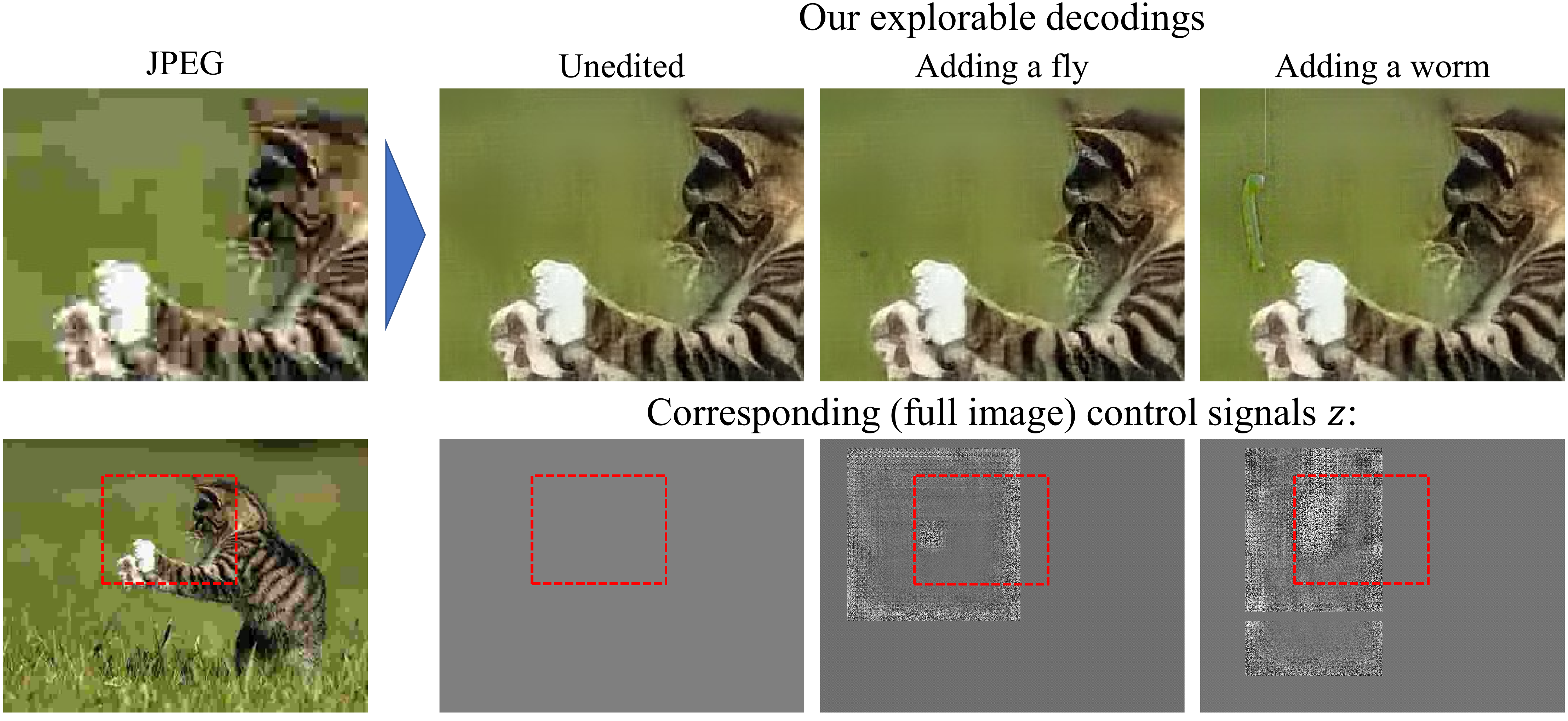}
\vspace{-10pt}\caption{\label{fig:kitten_and_fly_Zmap}\textbf{Possible explanations \& corresponding $\bf{z}$ signals.} Top row: zoom-in on the JPEG pipeline decoding (left) and three plausible (and perfectly consistent) alternative decodings by our method (repeated here for fluency from Fig.~\ref{fig:kitten_and_fly}). 
Bottom row: $z$ signals corresponding to our alternative decodings (entire image, zoomed region marked in red), displayed by reshaping each $1\times 1\times 64$ column in the $60\times 80\times 64$ $z$ signal to an $8\times 8$ block in a $480\times 640$ grayscale image (values were translated from $[-1,1]$ to $[0,1]$ for display purposes). While our pre-edited output (middle-left) corresponds to a constant $z$, to obtain the alternative appearances (middle-right and right), our imprinting tool's optimization process exploits the model's receptive field and modifies large regions of $z$ (limited to a rectangular window around the modified location), keeping non-modified parts of the image unchanged.
}
\end{figure*}

\begin{figure*}[t]
\centering
\includegraphics[width=\textwidth]{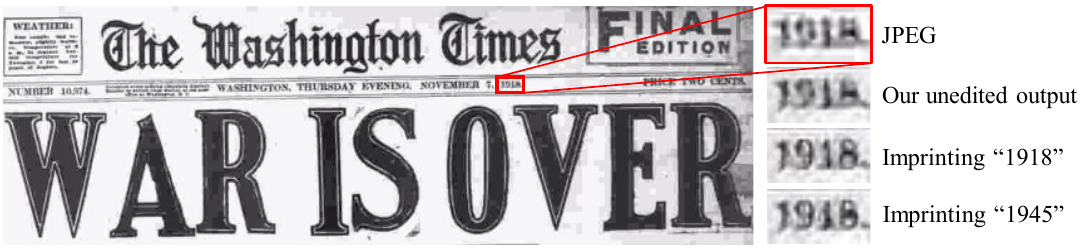}
\vspace{-10pt}\caption{\label{fig:newspaper}\textbf{Which war is over?} Using our framework to attempt imprinting years ``1918'' vs. ``1945'' yields a significantly better result for the former, suggesting this compressed archived newspaper dates back to the end of world war I.}
\end{figure*}
We next exemplify the exploration and editing process, and illustrate the effect of the quality factor (QF) on the available degrees of freedom. 


Figure~\ref{fig:editing_mole} shows the exploration process of Fig.~\ref{fig:applications}. Here, we attempted to imprint brown disks of varying radii on the unedited image. The top row shows the results of the first stage of the imprinting process, which projects the image with the naively placed brown disk onto the set of images that are consistent with the JPEG code. The second row shows the results of the second stage of the imprinting process, which seeks a control signal $z$ that causes the output of our decoding network to resemble the image produced in the first stage. In this example, the second stage is mostly responsible for smoothing some of the artifacts generated in the first stage.

Figure~\ref{fig:different_QFs} illustrates the effect of the QF on the space of images that are consistent with a given JPEG code. As can be seen, when using extreme compression with a QF of 5, we can produce consistent reconstructions with a wide range of mole sizes, from a very large mole on the left, to an almost vanished mole on the right. However, as the QF increases, the set of consistent solutions becomes smaller, making it impossible to imprint very large or very small disks.

Control signal $z$ has a non-local effect, due to the receptive field of our model. This is shown in Fig.~\ref{fig:kitten_and_fly_Zmap}, depicting different signals $z$ (bottom row) corresponding to different consistent decodings (top row) of a single compressed image. Finally, we present an additional exploration example in Fig.~\ref{fig:newspaper}, demonstrating a case of exploring corrupted text.
\section{Validating our Alternative Modeling of Chroma Subsampling}
In an effort to produce higher quality reconstruction of the chroma information, we wish to concatenate the reconstructed luminance information $\hat{x}^Y$ to the input of our chroma reconstruction model. However, this requires handling the dimensions inconsistency between the full-resolution luminance channel and the subsampled chroma channels, which we do through introducing an alternative modeling of the JPEG chroma subsampling process, as we explain in Sec.~\ref{sec:subsampling_remodeling}.

To validate this alternative modeling, we looked at the differences (calculated after going back to the pixel domain) between images undergoing the following original vs. alternative subsampling processes:
\begin{enumerate}
    \item \emph{``4:2:0'' JPEG pipeline:} Subsampling chroma channels by a factor of $2$ in both axes $\rightarrow$ Computing DCT coefficients for each $8\times8$ pixels block $\rightarrow$ Right and bottom zero-padding each coefficients block to $16\times16$ $\rightarrow$ Returning to pixel domain by computing inverse DCT for each $16\times16$ block.
    \item \emph{Our alternative pipeline:} Computing DCT coefficients for each $16\times16$ pixels block $\rightarrow$ Setting each block's $3$ lower-right quadrants to $0$, leaving unchanged the $8\times8$ upper left quadrant coefficients that correspond to low-frequency content $\rightarrow$ Returning to pixel domain by computing inverse DCT for each $16\times16$ block.
\end{enumerate}
Note that we did not perform any quantization step in either of the alternatives, as we were only interested in the isolated effect of remodeling the subsampling pipeline.

We computed the differences between the two alternatives using the RGB color representation, after concatenating back the non-altered luminance channel ($Y$) in both alternatives. We experimented using $100$ images from the BSD-100 dataset \cite{Martin2001bsd}, and found that the average root mean square error (RMSE) was a negligible $0.009137$ gray levels (corresponding to a PSNR of $88.9$dB). This certifies our decision to use the alternative modeling, which allows us to make our chroma reconstruction network aware of the corresponding luminance channel, by concatenating it to the network's input.
\section{Full Training Details}
We train our model on 1.15M images from the ImageNet training set \cite{russakovsky2015imagenet}, using batches of $16$ images each. We use an Adam optimizer, with learning rates of $0.0001$ and $0.00001$ for the initialization and consecutive training phases, respectively, and set $\beta_1=0.9$ and $\beta_2=0.999$ for both generator and critic networks. After \textapprox6 initialization phase epochs, we set $\lambda_{\text{Range}}$ and $\lambda_{\text{Map}}$ from Eq.~\eqref{eq:total_loss} to $200$ and $0.1$ respectively, and train for additional \textapprox12 epochs, performing $10$ critic iterations for every generator iteration.

For using ${\cal L}_{\text{Map}}$ in the latter phase, we feed each batch of images twice, differing by the input control signals $z\in[-1,1]^{m\times n\times 64}$. We first feed a per-channel constant $z$, uniformly sampled from $[1,-1]$, using only ${\cal L}_{\text{Range}}$ and ${\cal L}_{\text{Adv}}$ in \eqref{eq:total_loss}. We then perform $10$ minimization iterations over ${\cal L}_\text{Map}=\min_z\|\psi(x_Q,z)-x\|_1$, and feed the same image batch with the resulting $z$, this time using the entire cost function in \eqref{eq:total_loss}.

To create compressed image input codes, we compress the GT training images utilizing a quantization interval matrix \mbox{$M=\text{QF}\cdot Q_{\text{baseline}}/5000$}, where QF is independently sampled from a uniform distribution over $[5,49]$ for each image\footnote{We omit the upper QF range of $[50,100]$ when demonstrating our approach, as these higher QF values induce lower data loss, leaving less room for exploration.}, and $Q_{\text{baseline}}$ is the example baseline table in the JPEG standard \cite{wallace1992jpeg}. We use $N_\ell=10$ layers for both generator and critic models, where convolution operations utilize $3\times 3$ spatial kernels with $320$ or $160$ output channels for all layers but the last, in the luminance or chroma networks, respectively. We employ a conditional critic, which means we concatenate the generator's input $x_Q$ to our critic's input, as we find it accelerates training convergence.

\fi

\end{document}